I give permission for public access to my thesis and for copying to be done at the discretion of the archives' librarian and/or the College library.

_________________________________                    ______________________

Signature                                            Date

*BDELLOVIBRIO BACTERIOVORUS* PREDATION IN

DUAL-SPECIES BIOFILMS OF *E. COLI* PREY AND *M. LUTEUS* DECOYS

By

Phuong Binh Vo

A thesis submitted to the faculty of Mount Holyoke College in partial

fulfillment of the requirements of the degree of Bachelor of Arts with honor

Program in Biochemistry

Mount Holyoke College

South Hadley, MA

May 2010

This paper was prepared

under the direction of

Dr. Megan Nunez for 10 credits.

***For my family***
*Who gives me endless love,*
*Encourages me to shoot for the moon,*
*And reminds me to smile.*

# ACKNOWLEDGEMENTS

The support of my advisor, professors, family and friends was the key for the successful completion of this thesis.

I would like to express my deepest appreciation to my academic and thesis advisor, Professor Megan Nunez, for her generous guidance and inspiration. Her energy and enthusiasm in research had motivated me since the first day I joined her lab.

I am thankful to my thesis committee, Professor Jeff Knight and Professor Himali Jayathilake for being extremely understanding and supportive. Thank you to Marian Rice for technical support since the first day I started my research, and to my post doc, Anne Murdaugh for her generous help and advice.

My special thanks would go to my family for their continued love and support, financially and emotionally. They have given me the opportunity to study in the United States. They have been by my side no matter what happens. Without them, I would never go this far.

I thank my friends from all over the world, the chemistry lounge "regulars" and my lab partners who always cheer me up and listen to me, especially during my last semester. My special thanks would go to He Xu '12 for her collaboration during Spring '09 and Summer '09 and for Figures 11, 16,18 c and d, and 27 in this thesis. Thank you Mount Holyoke College for the research opportunities and most importantly, for being my second home.

Finally, I have to thank the National Science Foundation and Howard Hughes Medical Institute for making my research possible.

# TABLE OF CONTENTS

# LIST OF FIGURES

# ABSTRACT

Biofilms are matrix-enclosed microbial communities that grow at interfaces. They are highly robust and exhibit significant phenotypic changes that render them resistant to many antibacterial agents that can kill their free-swimming counterparts. Researchers have tried to find an effective, alternative and bio-friendly way to eliminate biofilms. Previous investigations in this group demonstrated that *Bdellovibrio bacteriovorus,* a small gram-negative predatory bacterium that consumes other gram-negative bacteria, could eventually eradicate a single-species *E. coli* biofilm in some conditions. These results suggest the potential value of *B. bacteriovorus* in biofilm eradication in industrial, medical and environmental contexts.

Biofilms in the environment can consist of either a single or multiple microbial species including both gram-negative bacteria and gram-positive bacteria. Here we investigated the potential of *Bdellovibrio bacteriovorus* to interact with and remove multi-species biofilms, specifically dual-species biofilms of gram-negative *E. coli* prey and gram-positive *M. luteus* decoy at interfaces betwee is more, less, or equally susceptible to *B. bacteriovorus* attack compared to their single-species counterparts.

Different research methods including bacterial culture, cell counting, crystal violet staining, gram staining, optical microscopy, scanning electron microscopy, and atomic force microscopy were explored to gain an insight into *Bdellovibrio's* interaction with biofilms in a macroscale and microscale. Our experiments showed that in a biofilm of *E. coli* gram-negative prey and *M. luteus* gram-positive decoy, *M. luteus* tend to form clusters in a columnar fashion and mostly grow on top of *E. coli* cells. However, with the presence of *B. bacteriovorus, B. bacteriovorus* not only consumes *E. coli* but also weakens the attachment of *M. luteus* to the solid surface, rendering the biofilms susceptible to removal. *B. bacteriovorus* controls not only prey but also decoy bacterial populations in the surrounding media, the latter probably via competition for nutrients. These experiments encourage us to consider how *B. bacteriovorus* might be used to control biofilms in the environment.

# CHAPTER 1: INTRODUCTION

## I. Biofilms

Biofilms are complex microbial communities that grow at interfaces, often at solid-liquid surfaces. Biofilms are usually found on solid substrates submerged in or exposed to some aqueous solution, although they can form as floating mats on liquid surfaces and also on the surface of leaves, particularly in high humidity climates. They can consist of a single microbial species or multiple microbial species, including different bacteria, algae and yeast.

Biofilm formation occurs by at least three different mechanisms (O' Toole *et al.*, 2000). In one mechanism, type IV pili-mediated twitching motility encourages surface aggregation. Alternatively, attached cells spread outward and upward by binary division to form cell clusters, or cells are recruited from the bulk fluid to form of biofilms. The relative contribution of these mechanisms depends on the organisms, the nature of the surface, and the physical – chemical condition of the environment. The twitching motility, growth rate, cell signaling, exopolysaccharide production, and the physical growth environment all play a significant role in the biofilm structure (Stoodley *et al.*, 2002).

### *a. Biofilm formation*

Differentiation in biofilm development has been explored since the 1980's. O' Toole and colleagues noted that the biofilm's structure undergoes a series of physical changes over time (O'Toole *et al*., 2000). Biofilm formation includes three steps: movement of planktonic cells in liquid, irreversible adhesion to a surface of biofilm, and biofilm maturation.

In the free-floating or planktonic stage, bacteria encounter a submerged surface and within minutes can become attached. They begin to produce extracellular polymeric substances (EPS) and to colonize the surface. These cells are capable of independent movement by twitching or gliding, and are not yet strongly adhered to the surface. Some might leave the surface for the planktonic lifestyle during this period of reversible adhesion.

During irreversible formation, the bacteria demonstrate several behaviors including rolling, creeping, and aggregate formation before secreting EPS and adhering strongly to the surface. EPS production mediates the transition from a weak interaction of cells to a permanent bonding between cells and the surface, thus allowing the emerging biofilm community to develop a complex, three-dimensional structure that is influenced by a variety of environmental factors. In addition to the EPS production, O' Toole *et al.* suggest that interactions of bacteria with one another at a surface, forming groups of cells, also help to strengthen the degree of attachment to the surface.

Gerke *et al*. (1998) also showed that adherent cells produce a protein called Polysaccharide Intercellular Adhesin that bonds the cells together and facilitates the formation of these microcolonies and the maturation of biofims. Mature biofilm communities can develop within hours.

The biofilm maturation process includes the generation of complex architecture, channels and pores, and a redistribution of bacteria swimming away (O' Toole *et al.,* 2000). As biofilms mature, they develop the basic microcolony and water channel architecture. Many cells change their physiological processes in response to conditions in their niches. Some microcolonies may detach from the surface or give rise to planktonic revertants that swim away from the matrix-enclosed structures. This activity leads to hollow remnants of microcolonies or empty spaces that become parts of the water channel. Allison *et al*. (1998) suggested that starvation accounts for the detachment that allows bacteria to search for nutrient-rich habitats. In addition, cell density may also trigger the release of degradative enzymes that allows bacteria to disperse from the matrix-enclosed environment when cell density reaches a high level in biofilm formation. However, the exact mechanism of how cells detach from a biofilm is still unknown. The dispersing biofilm cells revert to the planktonic lifestyle, allowing the biofilm developmental life cycle to come full circle (Figure 1).

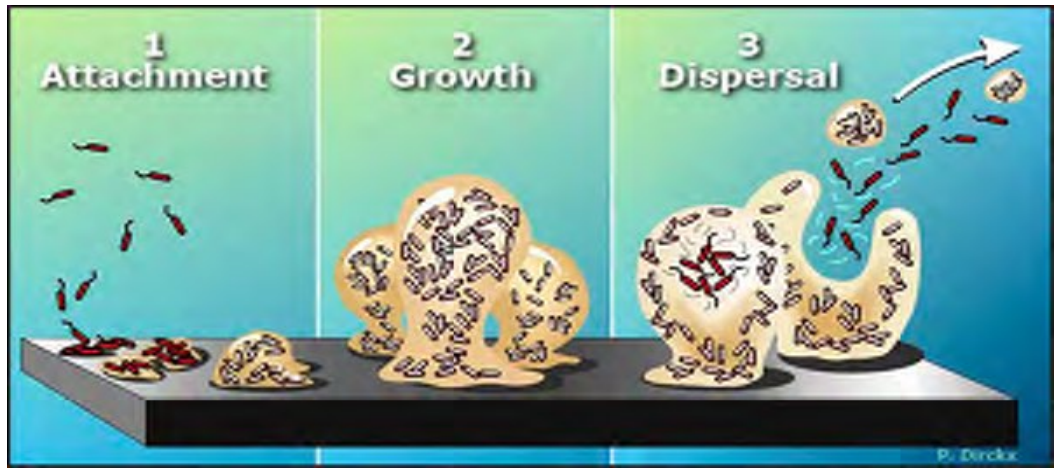

Figure 1: Three stages of biofilm development. (1) The free-swimming/
planktonic cells encounter a surface and initiate the biofilm growth. (2) The
biofilm mass grows from a weak attachment to a strong permanent attachment
to the surface. (3) When the biofilm matures, dispersal occurs. Some cells
detach from the biofilm and become free-swimming. Figure borrowed from
http://biofilmbook.hypertextbookshop.com/

### *b. Biofilms are harmful and hard to destroy*

Bacteria in biofilms are less mobile and more adhesive than their relatives. They stick together to form a complex community and carry out different roles. Often, biofilms are harmful to industry, the environment, and human health (Costerton *et al.,* 1999). For example, anaerobic bacteria in biofilms reduce sulfur to hydrogen sulfide to corrode pipes; aerobic bacteria use oxidation to corrode metal. On computer chips, biofilms serve as conductors to interfere with electronic signals. More than half of the infectious diseases caused by *Pseudomonas aeruginosa, Escherichia coli, Vibrio cholerae,* and other bacteria involve biofilms (Potera, 1996).

Biofilms are resistant to current modes of removal such as corrosive chemicals, bacteriophage, and antibiotics or immune cells (Watinick and Kolter, 2000). Therefore, biofilms are robust, diverse, and hard to destroy. Various techniques have been performed to manage and eliminate biofilms, such as chemical treatments, heat, and cleaning regimens. Recently, there has been interest in finding bio-friendly agents to eradicate biofilms. One organism that has been found to have a potential use against biofilms is the *Bdellovibrio* bacterium (Nunez *et al*., 2005).

## *II. Bdellovibrio bacteriovorus*

*Bdellovibrio bacteriovorus* was discovered in 1962 (Stolp and Petzhold, 1962). While trying to isolate bacteriophage from soil samples, Stolp and Petzhold noticed a number of odd plaques. These plaques took days to develop and continued to grow in size for more than a week. Using light microscopy, they observed small, highly motile and spiral shaped cells. Those cells are *Bdellovibrio* (Stolp, 1973)

*B.bacteriovorus* is a small (0.2-0.5 µm x 0.5-2.5µm), gram-negative bacterium that consumes other gram-negative bacteria. (Ruby *et al.,* 1991). Despite their small size, *Bdellovibrio* swim at high speed (from ~ 35 µm/ sec to ~160 µm/sec) with motility generated by the rotation of a single long flagellum that is polar and sheathed.

*Bdellovibrio*'s life cycle exhibits two major phases: a free-swimming attack phase and a growth phase (Figure 2). During a free-swimming attack phase, *Bdellovibrio* swims around to find its target, which is a gram-negative bacterium (Rittenberg, 1983). When the right target is encountered, *Bdellovibrio* penetrates the prey cell walls and establishes itself within the prey cell periplasm. The cell with *Bdellovibrio* growing inside rounds up and becomes a "bdelloplast." *Bdellovibrio* inside takes up the prey cell nutrients to grow and divide. When the cell nutrients are depleted, the bdelloplast is lysed, and the progeny are released into the environment to start a new cycle.

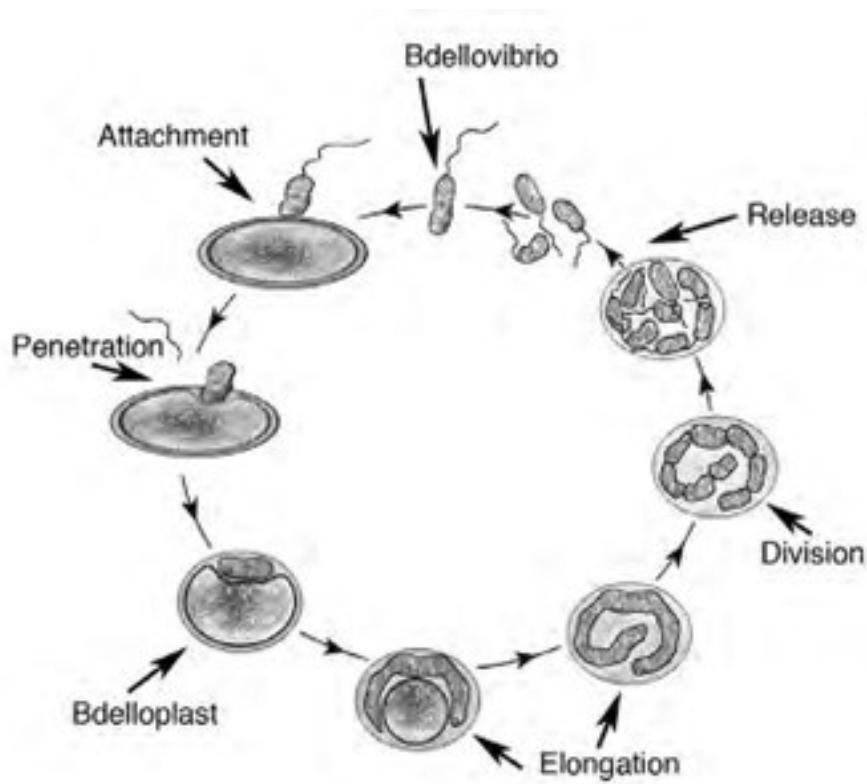

Figure 2: *Bdellovibrio* life cyle diagram. *Bdellovibrio* exhibits a 2-phase life cycle of a free swimming stage and a growth stage. After the attachment and penetration into the cell (top left), the cell rounds up with *Bdellovibrio* inside to form a bdelloplast (bottom left). The *Bdellovibrio* cell elongates, divides and releases progeny back into the water or soil environment (right). Figure borrowed from Volle *et al.*, 2008.

### a. Attack phase

During the attack phase, *Bdellovibrio* swim around to identify the prey. Thomashow and Rittenberg, (1985) observed that *Bdellovibrio*'s flagella obtain a characteristic complicated waveform. The complex waveform of the flagellar filament suggested that it consists a lot of protein. Thomashow and Rittenberg found that the flagellum is made up of six different flagellar filament proteins, but interestingly, only one of these is required for flagellar assembly, motility and efficient predation. This special feature allows *Bdellovibrio* to achieve high speed and helps the filament to be dispensable when needed (Lambert *et al*., 2006).

There seems to be a brief recognition period for *Bdellovibrio* to identify its prey after a collision with another cell (Shilo, 1969). Initially, the attachment to a cell surface is reversible. *Bdellovibrio* is still able to swim away a few seconds after recognizing that the cell is not a right target (gram-positive bacteria). When a gram-negative bacterium is encountered, *Bdellovibrio* cell becomes committed to invasion. The whole process usually takes around 5 – 10 minutes. *Bdellovibrio* drops its flagellum. It has been hypothesized that *Bdellovibrio* may adhere to the cell surface using pilus-like fibre structure expressed on its penetration pole.

The collision between the *Bdellovibrio* and the prey results in the flattening of the outer membrane of the prey cell. The tight coupling of *Bdellovibrio* to the prey weakens this region of the prey's cell wall, thus rendering an area of the wall susceptible to osmotic forces and producing a swelling budge in the prey cell. A pore on the prey cell wall is also created to allow *Bdellovibrio* to penetrate into the cell (Abram *et al.*, 1974). Once *Bdellovibrio* is inside the prey cell, the pore in the prey cell wall is resealed. At this point, *Bdellovibrio* is ready to move to the growth phase (Thomashow and Rittenberg, 1978).

### b. Growth phase

*Bdellovibrio* uses the first 30 minutes of the growth phase to prepare for cell multiplication. Inside the periplasm, *Bdellovibrio* establishes itself within the periplasm. In early work, Burnham *et al.* (1968) observed that *Bdellovibrio* occupies an invagination of the cytoplasmic membrane. *Bdellovibrio* does not necessarily need to occupy the whole space between the membrane and the cell wall in the prey cell (Scherff *et al.*, 1966).

The chemical modification of the prey's peptidoglycan results in the rounding up the prey cell. The size of *Bdellovibrio* inside the prey cell depends on which stage of the life cycle it is in. Starr and Baigent (1966) observed that *Bdellovibrio* in the prey cell, when fully nourished, may reach up to $3 - 4$ times as big as its original size. The longer *Bdellovibrio* stays inside the prey cell, the bigger it becomes. Statistically, some 3-6 *Bdellovibrio* progeny are produced from a single *E. coli* prey cell. More *Bdellovibrio* progeny are generated from bigger prey cells.

At 45 minutes into the growth phase, *Bdellovibrio* starts DNA replication. The *Bdellovibrio* directs a degradation of the prey cell's macromolecules for its biosynthesis and reproduction. Matin and Rittenberg (1972) observed the secretion of endonucleases and exonucleases of *Bdellovibrio* to cut the prey DNA. The prey deoxyribonucleotides are incorporated into the *Bdellovibrio*'s DNA until the *Bdellovibrio*'s genome replication is complete. At this point, the cytoplasm of the prey cell is very disorganized and severely damaged.

Together with the degradation of the prey macromolecules, the synthesis of predator ATP occurs concurrently with formation of a long filamentous cell as well as the separation of the cell into several progeny. The progeny become flagellate and remain inside the prey cell until the whole cell contents are depleted. Finally, the bdelloplast is lysed and the progeny are released into the environment. The new progeny become free-swimming attack phase cells, ready to invade their targets (Varon and Shilo, 1968).

### c. Host-independent Bdellovibrio

Although *Bdellovibrio* exhibit a host – dependent (HD) lifestyle, *Bdellovibrio* can also exist in a host-independent (HI) lifestyle (Figure 3). While HD *Bdellovibrio* requires a high prey density, HI *Bdellovibrio* grows in high – nutrient environment and can grow ten times bigger than host - dependent *Bdellovibrio*. In addition to abundant nutrients, HI *Bdellovibrio* can be induced by heat shock, by the presence of growth initiation factors, and mutations at the host interaction *(hit)* locus (Cotter and Thomashow, 1992). The balance between HD and HI growth in natural environment is unknown, but observation has shown that there may be a switch in between these two life styles (Ferguson, Spain *et al.,* unpublished results)

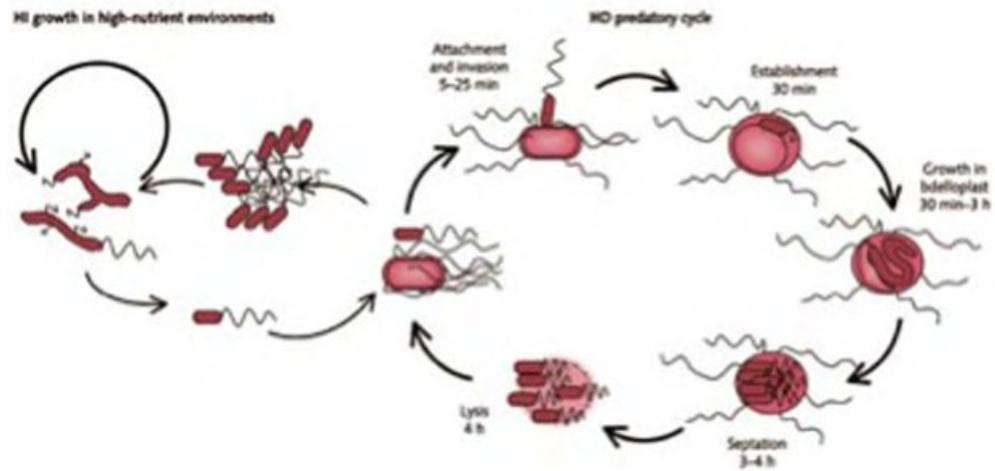

Figure 3: *Bdellovibrio* host-dependent (HD) and host-independent (HI) life style. HD *Bdellovibrio* requires the presence of gram-negative prey, while HI *Bdellovibrio* can grow with the absence of prey and take up nutrients from its environment. HI *Bdellovibrio* can be 10 times as big as a HD cell. Picture borrowed from Sockett, 2008.

**d.** *Metabolic pathways*

*Bdellovibrio* possesses enzymes of glycolysis as well as the pentose phosphate pathway. *Bdellovibrio* is capable of the aerobic metabolism of glycerol and synthesis of various Krebs cycle intermediates (Rittenberg and Shilo, 1970). Nonetheless, *Bdellovibrio* utilizes few carbohydrates as efficient sources of carbon and energy, instead deriving energy primarily from degradation of prey nucleic acids, proteins and lipids via aerobic pathways (Hespel *et al*., 1973).

*Bdellovibrio* is capable of synthesizing only 11 of 20 amino acids needed from intermediates of energy metabolism, thus the uptake of the prey amino acids is important to facilitate the rest of the amino acid synthesis of *Bdellovibrio* (Rendulic *et al.,* 2004). After degrading the prey macromolecules, they re-synthesize their own from the constituent nucleotides, sugars, or amino acids rather than directly transporting and using complex metabolic intermediates of the prey.

**e.** *Genetics of Bdellovibrio*

The sequence of the *Bdellovibrio* genome was published in 2004 (Rendulic *et al.,* 2004). It revealed that *Bdellovibrio*'s genome consists of a large number of genes compared to its small size. There are $3.8 \times 10^6$ base pairs (bp) on a single chromosome, which codes for 3,500 proteins.

Analysis of the genome sequence confirmed many previous results and hypotheses about *Bdellovibrio*. Scientists proposed several hypotheses for the adhesion and prey recognition of *Bdellovibrio* when the *Bdellovibrio* temporarily attaches to the prey. Huang and Starr, (1973) suggested that there must be some adhesion gene or some kind of protein interaction between the predator and its prey. Passive protein – protein interaction or LPS – LPS interactions between the outer membrane components of the predator and the prey were also suggested. In fact, when the genome of *Bdellovibrio* was elucidated, genes for adhesion were found. In particular, clusters of *pil* genes were found on the *Bdellovbrio* chromosome that could not be ascribed to other functions (Rendulic *et al.*, 2004).

Before the invasion of the prey cell, *Bdellovibrio* generates a small opening in the prey cell's outer membrane and peptidoglycan layer (Shilo, 1969). In order to do this without doing any excessive damage to the prey, *Bdellovibrio* use a mixture of hydrolytic enzymes to enter and digest the prey cell. Genes proteases including serine, cysteine, and aspartate proteases, as well as metal – dependent proteases, were found in the genome and are proposed to be involved in this process. *Bdellovibrio* secretes glycanases to solubilize the prey peptidoglycan early in invasion and these multiple glycanases were also found in the genome (Rendulic *et al.*, 2004).

Huang and Starr, (1973) suggested that there should be some enzymatic activity to facilitate the rounding up of the bdelloplasts and protect the bdellovibrio from oxidative damage by free radicals in the periplasm of the prey. It was found later that the hydrolysis and glycanse enzyme activity is deactived once *Bdellovibrio* has entered the periplasm and weakened the rigid structures of the cell wall (Thomashow and Rittenberg, 1978). More enzymes such as deoxyribonucleases, ribonucleases and lipases are found in *Bdellovibrio*'s genome to assist metabolic processes. These enzymes are abundant in the *Bdellovibrio* genome. (Rendulic *et al*., 2004)

Interestingly, very recent proteomic studies of the *Bdellovibrio* genome has begun to reveal how *Bdellovibrio* genes are expressed temporally during the growth phase, allowing *Bdellovibrio* to selectively utilize proteins and genetic information (Lambert *et al.,* 2010). These studies have revealed novel information too, such as that chaperones are expressed during bdelloplast phase to assist the non-covalent folding or re-folding and the assembly or disassembly of *Bdellovibrio* proteins during the protein synthesis.

### f. The habitat of Bdellovibrio

Bdellovibrios are found in various aerobic environments, such as soil or water.  When living in an aquatic environment, *Bdellovibrio* exhibits a preference for submerged surfaces. They can also be seen in fresh and brackish water, sewage, water reservoirs, and seawater. Due to the requirement for prey, *Bdellovibrio*s have been associated with biofilms (Williams *et al*., 1995). Biofilms offer good conditions for *Bdellovibrio*'s survival. It is suggested that in biofilm, *Bdellovibrio*s benefit from higher prey density.

*Bdellovibrio* interacts with abiotic environments as well as biotic environments. Baer *et al.,* (1994) observed that bdellovibrios in pure suspensions are able to adhere directly to sterile plastic and glass surfaces. Williams and his colleagues discovered that the nature of the surface also plays an important role in *Bdellovibrio* growth and survival (Williams *et al*., 1995). Surfaces in aquatic environments provide the predator with essential nutrients for growth, which enhances the survival of *Bdellovibrio* under extreme environmental conditions. The nature of the environment influences *Bdellovibrio* population. The cell density of *Bdellovibrio* on surfaces depends on the nature of the surface: a rough surface gives rise to higher numbers of bdellovibrio cells compared to other surfaces such as metal. The longer time that bdellovibrio cells have to interact with a submerged surface, the greater numbers of predators that appear on this surface biofilm. When a large number of predators associate rapidly with the surfaces, physical forces cannot easily displace bdellovibrios from their environments. The population of *Bdellovibrio* continues to survive and multiply as long as it is attached to a submerged surface and provided by a high prey density.

## III. Experimental questions

Previous research has shown that *Bdellovibrio* can eliminate the *E. coli* in a biofilm (Nunez *et al*., 2005). The Nunez lab has used prey-dependent *Bdellovibrio bacteriovorus* to destroy simple, single-species bacterial biofilms. However, biofilms in nature consist of multiple bacterial species, and the interaction of *Bdellovibrio* in a bi-species biofilm is still unknown. In this work, we have investigated how *Bdellovibrio* interacts with complex, multi-species bacterial biofilms at interfaces. To be specific, we are interested in how *Bdellovibrio* interacts with a biofilm that consists of both gram-negative prey and gram-positive decoys. With the presence of "inedible" gram-positive bacteria, are the multi-species biofilms more, less, or equally susceptible to *Bdellovibrio* attack?

In this lab, *E. coli* and *Micrococcus luteus* are used as a gram-negative prey and gram-positive decoy respectively. To investigate the interaction between these three types of bacteria, we have used a 24-well plate to mimic the natural biofilm formation. Three control groups were used to examine the properties of *E. coli* alone, *M. luteus* alone, and *E. coli-M. luteus* mixed communities. All three experimental groups involved *Bdellovibrio*, which specifically includes *Bdellovibrio - E. coli, Bdellovibrio - M. luteus*, and *Bdellovibrio - E. coli - M. luteus.*

We observed the interactions of these bacteria for five days, applying several techniques to both groups to examine the changes in biofilms over the time course. Cell culture and cell counting were performed to measure the cell density, while crystal violet staining, gram staining, and light microscopy were used to classify bacteria on the coverslip and also to observe the change of biofilms over time on a macroscopic level. Atomic force microscopy and scanning electron microscopy allowed us observe the biofilms on the micrometer to nanometer scale.

Our experiment showed that the addition of *Bdellovibrio* yields a reduction in cell density of *E. coli*. In a three-way mixed biofilm, together with the decrease of *E. coli* cells, *M. luteus* attachment to the biofilm is also weakened. Therefore, our results indicate the potential uses for *Bdellovibrio* in eliminating harmful bacteria in biofilms.

## CHAPTER 2: MATERIALS AND METHODS

### I. Bacterial growth media

The following media were prepared using milliQ water and sterilized by autoclaving immediately.

| Media | Ingredients |
|---|---|
| Luria Broth (LB) | 1% (w/v) tryptone<br>0.5% yeast extract<br>0.5% NaCl |
| LB agar plates | 1% (w/v) tryptone<br>0.5% yeast extract<br>0.5% NaCl<br>1.5% agar |
| HEPES Metal (HM) Buffer | 1M HEPES pH 7.6<br>0.1% $CaCl_2$,<br>0.1% $MgCl_2$ |
| Nutrient Broth (NB) | 0.03% beef extract<br>0.01% yeast extract<br>0.05% Casamino acids |
| Dilute Nutrient Broth (DNB) + metal | 0.03% beef extract<br>0.05% peptone<br>0.01% yeast extract<br>0.05% Casamino acids<br>Added after autoclaving:<br>  1.0 mM $CaCl_2$<br>  0.1 mM $MgCl_2$ |

## II. Cells

*Bdellovirbio bacteriovorus* 109J was obtained from the American Type Culture Collection (Manassas, Virginia, USA). *E. coli* strain ZK1056 was obtained from Professor Roberto Kolter (Havard Medical school), and *M. luteus* came from Professor Lynne McLandsborough (Univeristy of Massachusetts). *E. coli* strain ZK1056 and *Micrococcus luteus* were freshly cultured overnight in NB buffer prior to biofilm experiment.

## III. Biofilm model

A biofilm is a microbial community that grows on a surface or at an interface. In our lab, the 24-well plate with glass coverslip standing upright on a plastic stand was used as a biofilm model (Nunez *et al.* 2005). *E. coli* and *M. luteus* bacteria were grown freshly overnight to stationary phase, and 20 µL of this culture was added to 1.75 µL of DNB and metal buffer. Bdellovibrios were grown with *E. coli* until the culture was relatively clear of prey (3-6 days). This culture was filtered and 10 µL was added to the DNB. Coverslips were removed every day, from day 1 to day 5, to be stained and imaged under a light microscope and AFM. The growth medium in the well was diluted serially and plated on LB plates to determine the concentration of bacteria in a solution.

## IV. Serial dilution and cell counting

Cell counting was used to measure bacterial populations in a solution. In our lab, the number of *E. coli* and *M.luteus* cells in the well liquid was determined from day 0 to day 5. In serial dilution, the original bacterial inoculums were diluted in HM in a series of centrifuge tubes. Each tube will have one-tenth the number of microbial cells as the preceding tube.

$10^{-5}$ to $10^{-9}$ tubes were chosen to be plated on the petri dish. A 100 µL of solution of each tube was added to the plate and was spread uniformly over the surface of LB medium with a sterilized glass rod. After plating, the control group including *E. coli* and *M. luteus* were incubated at $30^0$C. The experimental group which involves *Bdellovibrio* was incubated at $37^0$C to kill *Bdellovibrio* so that the counted colonies represented only *E. coli* and/or *M. luteus*. *E. coli* appear small white colonies after a night of incubating, but it takes from two to three days for *M. luteus* to grow on a plate under these conditions. *M. luteus* once growing appears as yellow colonies.

## V. Crystal violet staining

Crystal violet (0.1% solution in water) was the primary stain used to detect whether bacteria were adhered to the glass coverslips. After one minute submerged in dye, the coverslips were washed with water. The biofilm stained as a purple line on the glass at the air-liquid interface, and becoming less purple towards the lower half of the coverslips. Since crystal violet is an amphiphic compound, it has an affinity for many amphiphilic biomolecules. It stains also debris, extracellular material, and all other contents that contain amphiphilic biomolcules. This method was used for qualitative detection of biofilm formation.

## VI. Gram staining

Gram staining procedure is a technique to sort bacteria into two broad categories, gram-negative and gram-positive. This technique is used in our lab to distinguish *E. coli* and *M. luteus*. A heat-fixed smear coverslip was stained with crystal violet for one minute before washing with water. After being washed with water, the coverslip was covered with iodine for another minute. When the iodine was washed off both types of bacteria appeared dark violet. The coverslip was decolorized with 95% Ethanol to remove the purple from the cells of gram-negative species but not gram-positive species. The coverslip was carefully rinsed off, and stained with safranin, a basic red dye, for another minute. The dye was washed off with water. After air drying, the coverslip was ready to be examined under a light microscope.

## VII. Light microscopy

After gram staining, a light microscope (Olympus BH2) was used to classify the types of bacteria and observe a coverslip in a macro scale. A range of lenses were used with different objective magnification, from 10x A10PL 0.25NA, 40x A40PL 0.65NA, to 100x A100PL 1.30 oil. Under highest magnification, 100x objective lens magnification, *M. luteus*, a gram-positive bacterium, appears as a purple round shape, while *E. coli*, a gram-negative bacterium, is in a pink rod shape. Images were captured by PixeLink Color camera with PLB681 CU model, then exported to Adobe Photoshop for final adjustment of dimensions and brightness.

The same microscope was used to view transparent specimens, in our case, the movement of native *Bdellovibrio* cells. After fresh culturing from stock for 5 days, *Bdellovibrio* was observed under phase contrast mode. The movement of significant numbers of bdellovibrio cells indicates that *Bdellovibrio* is ready for biofilm experiment.

## VIII. Atomic Forced Microscopy (AFM)

Coverslips were removed from the well plate, rinsed gently with water and left to air dry. The coverslips were fastened to stainless-steel AFM sample discs (Ted Pella Inc., Redding, CA) with double-sided adhesive tape. Bacteria were imaged immediately up to several hours after removal from the well plate by contact-mode AFM in air using a Digital Instruments Multimode SPM with a Nanoscope IV controller (Veeco, Santa Barbara, CA). Oxide-sharpened silicon nitride tips were used for good quality images. Images were flattened and exported to Adobe Photoshop for final adjustment of dimensions and brightness.

## IX. Scanning Electron Microscopy (SEM)

Coverslips were removed from the cultured well plate and move to an empty one to be treated before SEM imaging. The coverslips were rinsed briefly with PBS, then fixed with 0.1M Na Cacodylate buffer and post fixed with 1% Osmium. The Osmium would be removed by rinsing the coverslips with distilled water. The bacterial cells on the coverslips were dehydrated in each of a graded ethanol series of 50%, 70%, 80%, 90% and 100%, and critical point dried. The coverslips were gold coated and fastened to a stainless-steel SEM stud with carbon tape before imaging. Bacteria were imaged with FEI Quanta 200 SEM.  Images were exported to Adobe Photoshop for further adjustment of brightness if necessary.

# CHAPTER 3: RESULTS

In this experiment, we examined the interaction of *Bdellovibrio* with mixed *E. coli* and *M. luteus* biofilms. Cell counting, crystal violet staining, gram-staining, light microscopy, SEM, and AFM were used to observe the biofilms from a macroscale to a microscale for this experiment. All of the experiments were performed on both the control groups (without bdellovibrios) and the experimental groups (with bdellovibrios). The results of both groups were examined and compared to see the difference before and after bdellovibrios were added.

## I. Growing and counting *E. coli* cell population

In order to quantitatively measure the change of *E. coli* and *M. luteus* cell population over time, we developed conditions that yield stable *E. coli* biofilms in a 24-well plate. The initial predator/ prey/ decoy volume ratio was 1:2:2 in each well. After 24 hours, each group was plated on LB plates, and the number of colonies was recorded. Using back calculation, we obtained the cell counts of *E. coli* over the course of *Bdellovibrio* infection.

### a. *E. coli cell population*

Without the competition for nutrition and oxygen, *E. coli* alone reaches the largest and especially on the first and second day reaching a maximum growth of $5 \times 10^9$ cells/mL. The concentration of live *E. coli* cells starts to decrease by half each day between day 3 to day 5. The trend was consistent over four repeated trials of observation, which can be seen by the small error bars on each day of the cell counting data (Figure 4).

When other bacteria are added to the well, the *E. coli* cell density is much lower. In the case of *M. luteus* added in *E. coli* culture, maximum *E. coli* cell density decreases by 10 times. However, the drop in the cell population between days 3 and 5 is minimal.

When *Bdellovibrio* is involved, the *E. coli* cell density shares the same pattern of initial rapid increase followed by a slow decline of *E. coli*. To be specific, the number of *E. coli* in our experimental groups falls in the range of $10^9$ cells/mL during day 1-3, which is as much as that of *E. coli* in mixed *E. coli - M. luteus* liquid culture. The number of cells in the experimental groups also starts to decrease from day 3, but with a more dramatic drop.

In a mixed *Bdellovibrio – E. coli – M. luteus* cell cutures, the *E. coli* cell proliferation on the first two days is the same as in the *Bdellovibrio – E. coli* cultures. However, on day 4 and 5, the number of *E. coli* in a three-way mixture is 5-10 times less than that of a two-way mixed well liquid of *E. coli* and *Bdellovibrio*, and 100 times less than that of a mixed well liquid of *E. coli* and *M. luteus*. We noticed that with the presence of *Bdellovibrio*, the ending point (day 5) is lower than the starting point of day 0. With the absence of *Bdellovibrio*, the cell density on day 5 ends up 15 times higher than the starting level. In all cases, the initial amount of *E. coli* was the same.

The cell density of *E. coli*-only liquid culture shows the highest consistency throughout the whole experiment. In contrast, we measured more cell density variability in the two-way mixed and three-way mixed cell cultures, which is reflected in different lengths of error bars during the experiment period.

### b. *M. luteus cell population*

The cell density of *M. luteus* is about 10 to 100 times less than that of *E. coli* under the same conditions (Figure 5). Interestingly, *M. luteus* cell density shares the same pattern with *E. coli*. Specifically, the cell density in *M. luteus* reaches a maximum growth when it is alone. It is lower in *E. coli*-*M. luteus* mixed well liquid, and cell density is even lower when *Bdellovibrio* is involved. Together with the presence of *Bdellovibrio*, the presence of *E. coli* prey in the mixture is correlated the most dramatic drops in the number of *M. luteus* (Figure 5). While the plot of *M. luteus* in *E. coli* – *M. luteus* mixed well plate plateaus at $10^7 - 10^8$ cells/mL, the number of *M. luteus* cells with the presence of *Bdellovibrio* decreases notably over time to an overall concentration ten times less than *M. luteus* alone.

In the bacterial communities containing only *M. luteus*, *M. luteus* shows the most marked growth from on days 2 and 3. With the addition of *E. coli,* the *M. luteus* population drops by 5 times. The cells proliferate initially and diminish after day 2 by only 2 times, which forms a plateau shape. The cell count trend in *M. luteus* in *E. coli – M. luteus* mixed cultures is consistent with the *E. coli* cell density in the presence of *M. luteus.*

With the presence *Bdellovibrio*, the *M. luteus* cell growth stops after day 2 and starts to decrease dramatically on day 3 by 10 times. Interestingly, *M. luteus* mixed remains almost the same on day 3 and day 4, then drops to the same amount as *M. luteus* on the $5^{th}$ day (Figure 5).

In a bacterial culture of *Bdellovibrio – E. coli – M. luteus* mixed, the cell population does not thrive as much as the other cultures. The *M. luteus* cell density decreases from day 3 to day 4 in the three-way mixed culture. Without *Bdellovibrio*, the population of *M. luteus* on day 5 is roughly ten-times higher than the starting point of day 0. On the other hand, with the addition of *Bdellovibrio*, the number of *M. luteus* drops 5 times lower than the starting point of day 0.

The cell density of *M. luteus*-only culture shows the highest consistency throughout the whole experiment. In contrast, we measured more variability cell density in the two-way mixed and three-way mixed cell culture, which is reflected in different length of error bars during the experiment period.

## II.  Crystal violet staining

Crystal violet staining is used to detect the biofilm communities that adhere to the glass coverslip. The more dense the biofilm, the darker the color of the crystal violet stain. *E. coli* fresh liquid cultures were grown in 24-well plates containing upright, round 15mm sterile glass coverslips. *E. coli* biofilms formed densely at the air-liquid interface on coverslips half-submerged in growth medium.   The biofilms were stained with crystal violet. The blue color was observed in the liquid and most concentrated at the air-liquid interface. The air portion was less blue due to the lack of cells growth. In the absence of cells, the coverslip submerged in media was clear after staining (Figure 6).

Crystal violet staining was further used to see the overall pattern of biofilm density in the control group (Figure 7) and the experimental group (Figure 8) over the course of 5 days. The more cells a biofilm has, the darker the stain color. Based on our previous cell counting result, we hypothesized that the number of cells in the biofilms would initially increase and then drop from day 3, and with the addition of *Bdellovibrio*, the decrease in cell density would be more dramatic. The control group and the experimental group of cell cultures were removed from growth medium and stained with 0.1% crystal violet stain daily between day 1 to day 5.  Indeed, the staining color on all of the biofilms appeared darkest on day 2 and day 3, indicating the maximum growth of cells on these days. The color started to fade on day 4 and become very light on day 5, consistent with the decline in cell counts on the $5^{th}$ day
(Figure 8). Compared to the biofilms without *Bdellovibrio*, there was a subtle
decrease in the staining color of the experimental groups, but the decline was
not marked.

### III.   Imaging biofilms using light miscrosopy, SEM and AFM

Our crystal violet staining showed that there would be few or no cells
above the surface of the liquid in air, and the cell population was mostly
concentrated at the  air-liquid interface. In fact, by gram-staining and light
microscopy, we found that in all cases, a dense biofilm appeared at the
interface between liquid growth medium and air, while almost no bacteria
grew on the glass in the air (Figure 9). The sterile plastic stand that was placed
in the wells to hold the coverslips in an upright position was also examined
with crystal violet stain for cell adhesion. The stain was much lighter
compared to the glass coverslip, so fewer cells appeared to adhere to the
plastic cap (data not shown). Apparently, both *E. coli* and *M. luteus* cells form
a thicker and more robust biofilm on the glass surface than plastic surface.

In this experiment, we attempted to capture the interaction between
bdellovibrios, *E. coli,* and *M. luteus* by light microscopy, SEM, and AFM.
The biofilms grown on glass coverslips were imaged every 24 hours for the
change in macro-scale and micro-scale details. The change of *E. coli* cells and
*M. luteus* cells with and without the presence of *Bdellovibrio* in biofilms was

observed on a macro-scale using gram-staining and light microscope. Then with the SEM and AFM, we were able to observe the biofilms on the micrometer to nanometer scale.

### a. *E. coli* alone and *M. luteus* alone images

*E. coli* alone and *M. luteus* alone biofilms were imaged by light microscopy after gram-staining. Under these conditions, *E. coli* cells are pink and oblong (Figure 9). *M. luteus,* on the other hand, has a round, grape-like shape and smooth texture. Under light miscrosope, the *M. luteus* cells are stained purple (Figure 10). The darker the stain color, the denser the population in the cluster.

Our AFM and SEM studies showed that oblong *E. coli* tend to form a biofilm in a flat, two-dimensional fashion across on the glass surface (Figure 11 and 12). Although we found that occasionally *E. coli* do grow on top of each other and form a robust biofilm in SEM (data not shown), our AFM images showed that *E. coli* cells can only reach 200 – 400nm in height.

By SEM and AFM, we could see that *M. luteus* cells form clusters in a columnar fashion (Figures 13 and 14). Especially under AFM, the difference in height of *M. luteus* cells is shown as different colors from bright yellow to dark brown in a height image (Figure 13b). The higher the cells from the surface, the brighter the color. Some *M. luteus* cell clusters can grow on top of each other and form in a very high column, which shows in a bright-lit area of the AFM height image. The AFM z scale shows that the *M. luteus* alone biofilm can form structures up to 850 nm in height.

The cell counts of *E. coli* alone and *M. luteus* alone in the liquid around the biofilm were consistent with our microscopic pictures. Images of *E. coli* alone biofilms and *M. luteus* alone biofilms growth on glass coverslips were taken every day, from day 1 to day 5. We observed that these two single-species biofilms formed stable and dense biofilms during the first 3 days, but the biofilms began to degrade on day 4 and 5 (Figures 15 and 16).

### b.    Mixed E. coli – M. luteus biofilms

We examined the interaction between the round *M. luteus* and rod shaped *E. coli* in mixed bifoilms of these bacteria. *E. coli – M. luteus* mixed biofilms were gram-stained and observed under the light microscope (Figure 17). Round purple stained *M. luteus* grow in aggregates, often on top of the oblong pink stained *E. coli* monolayers. Due to this interaction, *E. coli* can only be found in the edges of *M. luteus* aggregates. *E. coli* can be seen on the first 3 days, but during later days, *E. coli* become more difficult to find on day 4 and 5, possibly because the *M. luteus* have grown to completely cover them.

After observing the changes in the biofilms at macro-level using the optical microscope, AFM and SEM were used to examine the changes in micro-level (Figures 18 and 19). Again, *M. luteus* dominates the surface in a mixed biofilm of *E. coli* and *M. luteus.* The oblong *E. coli* are adhered to the surface in contiguous sheets of cells, while the round and smooth *M. luteus* grows in a vertical fashion on top of each other and the *E. coli*. In our AFM pictures, *E. coli* cells are hardly to be seen and tend to sit underneath *M. luteus* cells, yet they can be found around the edges of *M. luteus*. Figure 18 is characteristic of a biofilm of *E. coli* and *M. luteus* under AFM. With a high resolution of the AFM, sticky extracellular secretions of the cells can clearly be seen in a small scan size (Figure 18a). In the light microscope, *E. coli* is out of focus (blurry) while *M. luteus* is in focus (clear) showing the huge height difference between these two cell types, but in the AFM we can measure heights and achieve good vertical resolution. With the addition of *E. coli* cells in *M. luteus* biofilms, the height of *M. luteus* is measured up to 1.5μm, while *E. coli* cells still lay flat on the surface, forming a monolayer around 200-400nm high (Figure 18b and d).

Using SEM, we also confirmed that *M. luteus* cell clusters grow on top of each other and on *E. coli* cells. *E. coli* are found underneath *M. luteus* cells and on the glass surface (Figure 19a). Occasionally, there is a well-mixed community of *E. coli* and *M. luteus* where *E. coli* mingles with the *M. luteus* and sometimes stay on top (Figure 19b), but generally the vertical stratification of species within the biofilm occurs.

### c. Mixed Bdellovibrio – E. coli biofilms

When *Bdellovibrio* is added to *E. coli* biofilms, the *E. coli* cell density is gradually eliminated. The microscopic pictures of *E. coli* cells in *Bdellovibrio – E. coli* biofilms agree qualitatively with the cell counting of *E. coli* in the liquid phase (Figure 20). Indeed, together with the decrease of *E. coli* well liquid cell counts, *E. coli* biofilms formed on the glass coverslips also decrease in thickness. On day 5 of the experiment, there are not many *E. coli* cells left (Figure 20e).

While *Bdellovibrio* cannot be seen under light microscope due to its small size and the low magnification of the light microscope, the predator can be seen under AFM and SEM. On day 4 and day 5, although a few *E. coli* are found, mostly well-fed bdellovibrios and some bdelloplasts are seen (data not shown). Under SEM, oblong *Bdellovibrio,* small in size (~ 0.5-1.0μm) with a characteristic flagellum, can be spotted in a mixture among the *E. coli* cells in a biofilm (Figure 21). Bdelloplasts can also be found in SEM. Bdelloplasts under SEM have a wrinkled surface with a *Bdellovibrio* growing inside. Due to the treatment with ethanol before imaging, the shape of some bdelloplasts gets distorted (Figure 22).

### d. *Mixed Bdellovibrio – M. luteus biofilms*

In a mixed biofilm of *Bdellovibrio* and "inedible" *M. luteus,* no *Bdellovibrio* were observed in the light microscope images due to the small size of *Bdellovibrio* and the low magnification of the light microscope. Only *M. luteus* is seen (Figure 23). *M. luteus* grows in aggregates and is concentrated at the solid – liquid interface. Compared to a single – species *M. luteus* biofilm, the biofilm in mixed *M. luteus – Bdellovibrio biofilms* has more void space, indicating the drop in *M. luteus* population when *Bdellovibrio* is involved (Figure 23). This finding correlates with our cell counting results. It indicates the elimination of *M. luteus* cells in a biofilm with *Bdellovibrio*.

While AFM has the resolution to image *Bdellovibrio*, the images of *Bdellovibrio – M. luteus* mixture so far have yielded up only *M. luteus* cells (Figure 24). Because *Bdellovibrio* also cannot be seen in the mixture of *M. luteus* and *Bdellovibrio*, it is hard to distinguish the difference between this mixed *Bdellovibrio – M. luteus* biofilm and *M. luteus* alone control biofilms (Figure 23). However, the height of *M. luteus* columns with the addition of *Bdellovibrio* is as high as *M. luteus* with *E. coli*. *M. luteus* cells are measured to be 1.5μm away from the surface.

### e.  Mixed Bdellovibrio – E. coli – M. luteus biofilms

Compared to *E. coli - M. luteus* biofilm, the cell density of *E. coli* and *M. luteus* in a three-way mixed biofilm drops even further. Gram-staining pictures of the three-species mixed biofilms as viewed under the light microscope show with more empty space, indicating the decrease in biofilm density in the presence of *Bdellovibrio* (Figure 25). The same biofilms were imaged by SEM, and as we expected, fewer cells are found in the presence of *Bdellovibrio* predator (Figure 26). This elimination makes the biofilm less dense with more void space.

We also noticed a phenotypic change of some *M. luteus* cells in the three-species biofilms. Although this observation was not found under AFM probably due to the small scan size, under light microscope and SEM, some *M. luteus* cells appear smaller and less round (Figure 25b, c and d). They tend to aggregate in units of two or four cells. Under SEM, the smaller and less round *M. luteus* cells appeared bright in color and on top of some *E. coli* cells (Figure 26). We have not seen the change in morphology in the other mixed biofilms, indicating that the presence of *Bdellovibrio* in mixed *E. coli* and *M. luteus* biofilms might have some effect on *M. luteus* morphology, thus leading to the change in the shape of some *M. luteus* cells.

However, the interaction between *M. luteus* and *E. coli* remains the same. *M. luteus* grow on top of *E. coli* with a biofilm depth of 1.5μm for *M. luteus*; *E. coli* lay flat on the surface with a height of 200-400 nm (Figure 27).

In the three-species biofilms, more *E. coli* are exposed to the surface, which correlates with the decrease in cell counts in *M. luteus* biofilms. The decline in both *E. coli* cells and *M. luteus* cells in biofilms correlate with our cell counting results, suggesting the ability of *Bdellovibrio* to both prey on *E. coli* and also to somehow reduce the *M. luteus* population.

The predation of *Bdellovibrio* on *E. coli* cells can be demonstrated by the formation of bdelloplasts. Using AFM and SEM, we were able to see the bdelloplasts in the mixed *Bdellovibrio – E. coli – M. luteus* biofilms. In SEM images, the round bdelloplasts are small in size (les than 1µm in diameter) and have a bdellovibrio cell growing inside, which appear brighter in color, emerging from the rest of the bdelloplast (Figure 28). Under AFM, round bdelloplasts can generally be identified by a smooth cell surface and the presence of a bdellovibrio growing inside (Nunez *et al.,* 2005). However under AFM, the similarity in appearance between *M. luteus* and bdelloplasts makes it harder for us to identify the bdelloplasts in the three-way mixed biofilm. The difficulty in imaging the bdelloplasts increases when AFM tip artifact is involved due to a great variability in height, and when *M. luteus* are densely packed and divide (Figure 29).

A large area of void space is also exposed when *Bdellovibrio* is introduced to prey-decoy biofilms (Figure 30). These experiments revealed the mechanism of how *Bdellovibrio* consume *E. coli* cells underneath *M. luteus* cell aggregates. Under light microscope, with the addition of *Bdellovibrio*, a small portion of *M. luteus* cells attaches to an *E. coli* cluster, but the majority of *M. luteus* cells seem to fall off, exposing the *E. coli* foundation underneath. This finding suggests that the presence of *Bdellovibrio* loosens the attachment between *M. luteus* and the solid surface, rendering this part of the biofilm susceptible to removal. The ability of *Bdellovibrio* to remove dual species biofilms is also confirmed by AFM and SEM pictures.

Under AFM, we can characterize numbers of *E. coli* cells that are exposed to the surface (Figure 31). The texture of these *E. coli* cells is not as smooth as the usual *E. coli* found in the previous AFM pictures. It is suggested that the number of *M. luteus* that used to stay on top of the *E. coli* fell off rendering the  *E. coli* foundation underneath available for *Bdellovibrio* predation. The corresponding height image shows that *M. luteus* cells can be as high as 4.0μm in this case.

SEM revealed the complex nature of a three-species biofilm (Figure 32). The biofilm contains large, tall aggregates of cells and EPS. Nevertheless, with *Bdellovibrio*, a big portion of *E. coli* is exposed in *M. luteus* clusters. The texture of this area is somewhat distorted and flattened, which suggests the *M. luteus* cells might have fallen off to render the *E. coli* part underneath for *Bdellovibrio* to attack. In a closer inspection, some bdellovibrio cells are found in the flat, exposed *E. coli* portion.

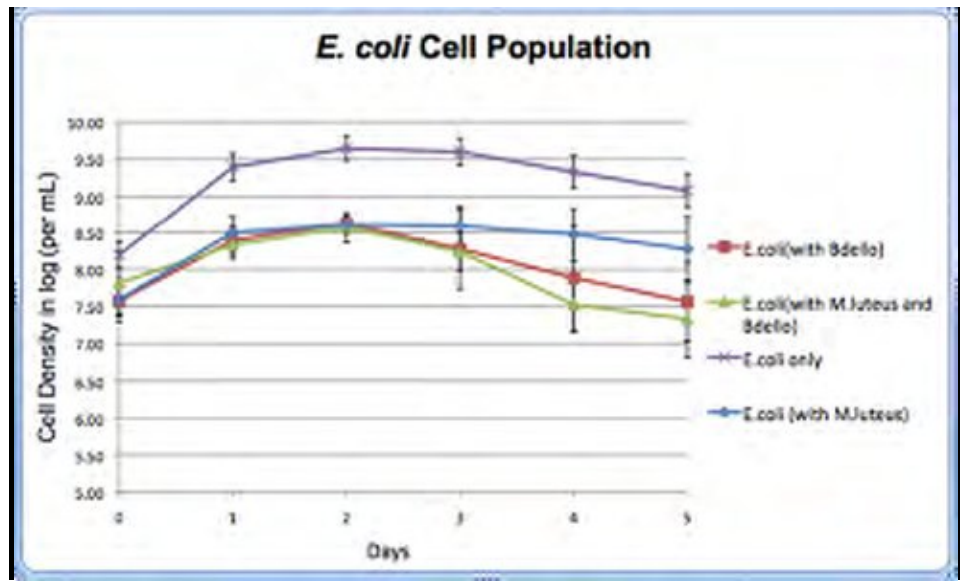

Figure 4.: Change in *E. coli* population density over the course of *Bdellovibrio* infection. The population size is expressed by the log of the number of cells per mL and is plotted against the time from the start date of the experiment. Without competition for nutrition and oxygen availability, *E. coli* alone shows the largest population. With the addition of *M. luteus* bacteria, the *E. coli* population is initially diminished due to competition for nutrients and oxygen. In addition to the lack of nutrition and oxygen, with the presence of *Bdellovibrio*, the predation further decreases the *E. coli* population.

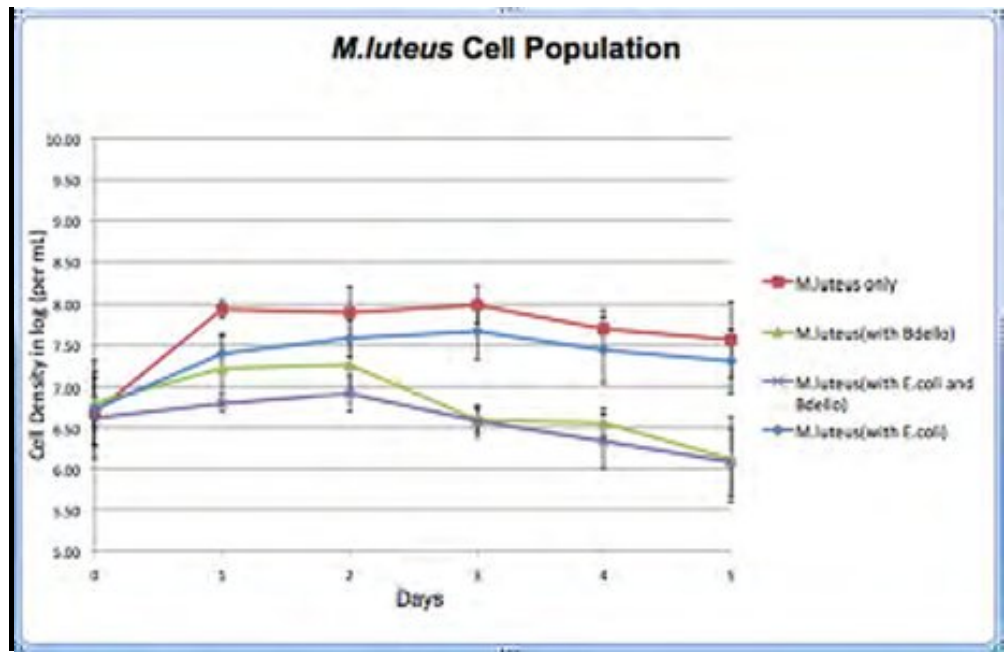

Figure 5: Change in *M. luteus* population density over the course of *Bdellovibrio* infection. The population size is expressed by the log of the number of cells per mL and is plotted against the time from the start date of the experiment. Although the cell population of *M. luteus* is ten-fold lower than that of *E. coli,* similar trends in population are observed for *M. luteus*. Even though *M. luteus* is not consumed by *Bdellovibrio*, with the presence of *Bdellovibrio*, the population is diminished due to nutrient and oxygen competition. As a decoy in the biofilm environment, *M. luteus* encourages *Bdellovibrio* predation on *E. coli*. *M. luteus* serves as a crucial factor of balancing the environment in a mixed biofilm.

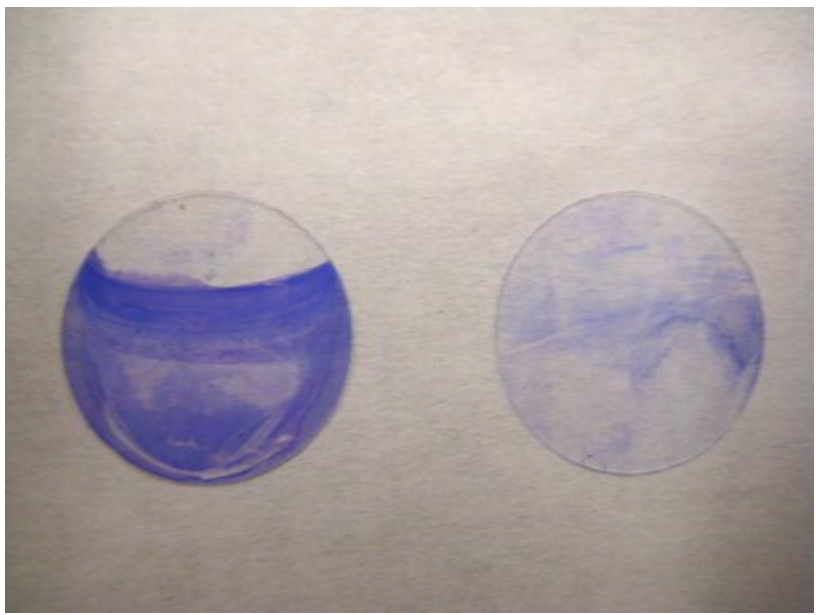

Figure 6: *E. coli* biofilm density on coverslips half-submerged in growth

medium. The biofilms were stained with crystal violet (left). Blue color was observed in the liquid phase but not in the air above. A coverslip placed in growth medium without cells (right) was clear after staining.

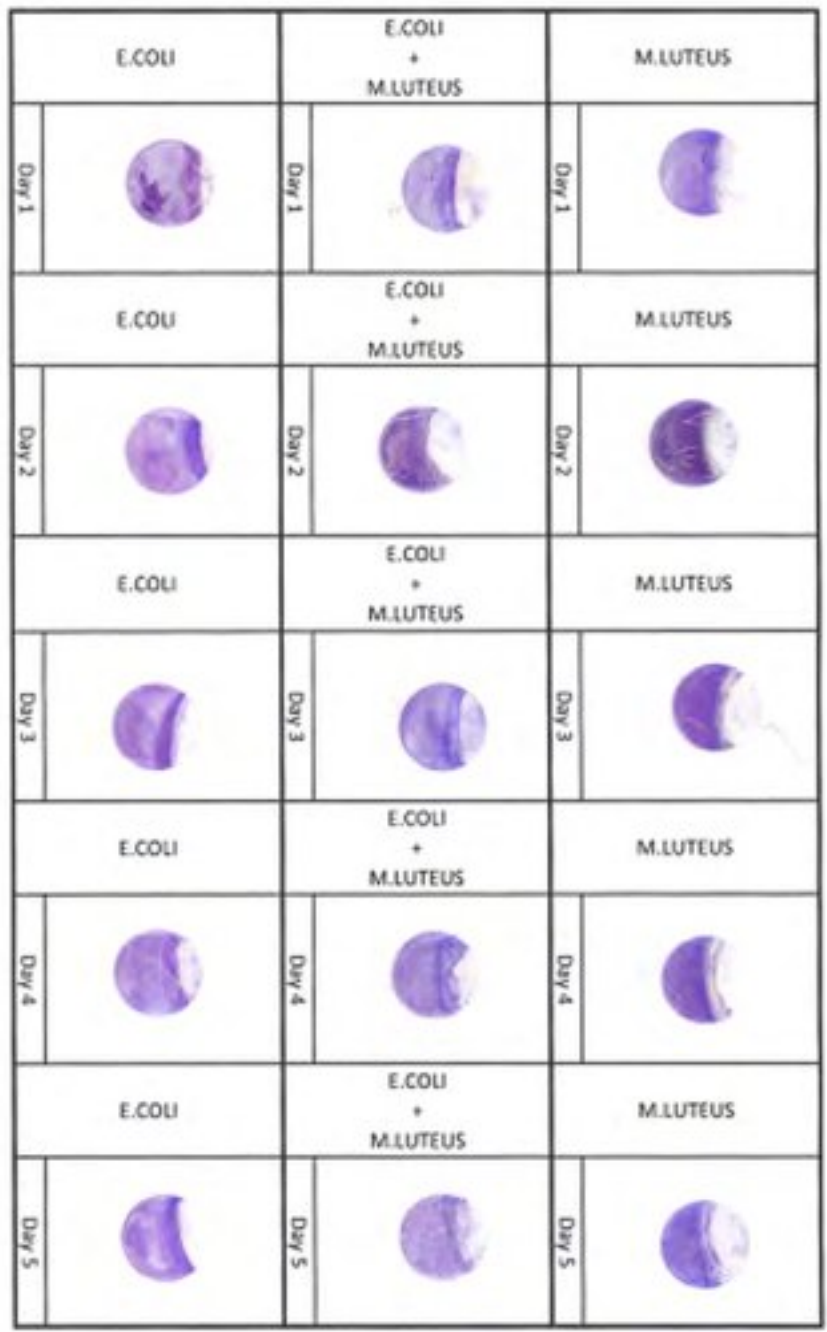

Figure 7: Biofilms of *M. luteus* alone (right), *E. coli* and *M. luteus* mixed

cultures (middle), and *E. coli* alone (left) were removed from growth medium and stained with 0.1% crystal violet stain daily between day 1 to day 5. The more dense the biofilm, the more purple is the coverslip. The biofilm density increases from day 1 to day 3, then decreases from day 4 to day 5.

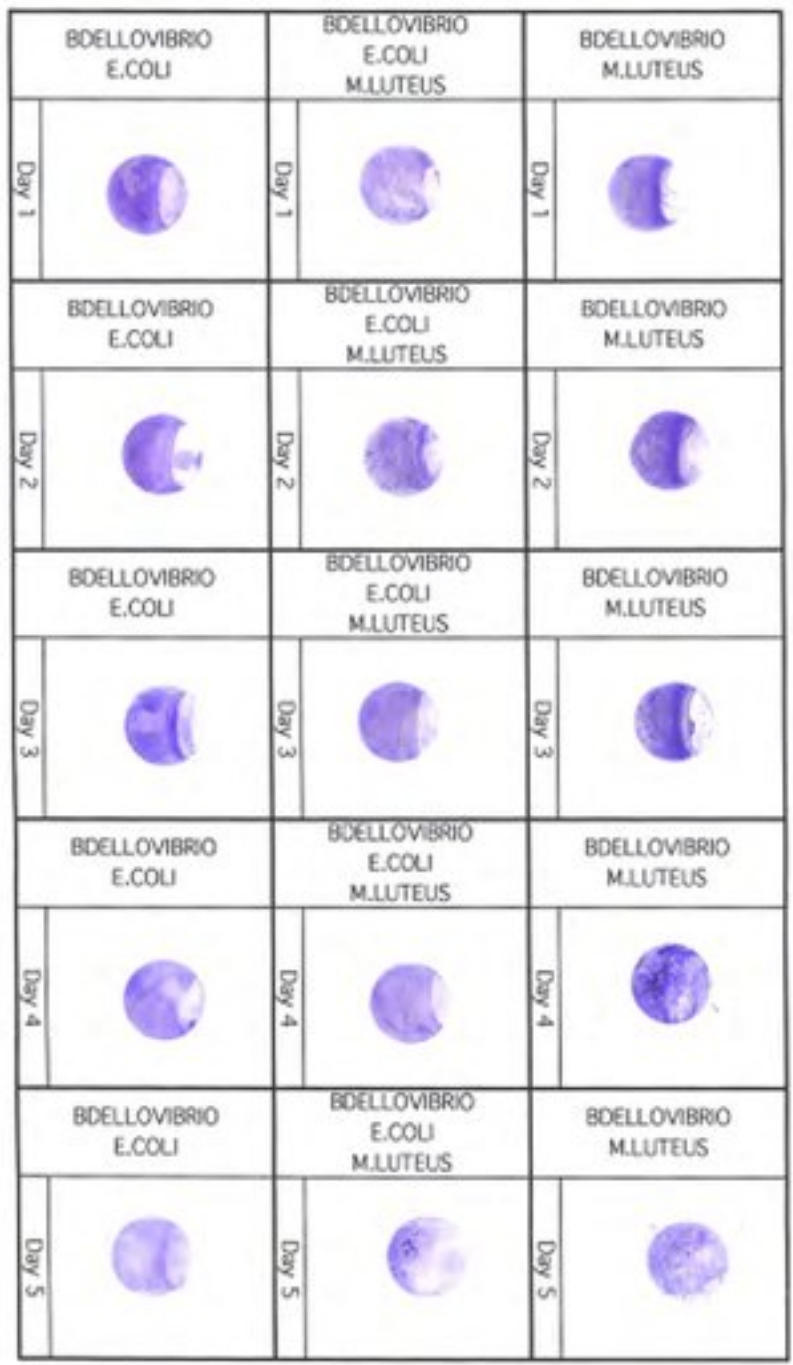

Figure 8: Biofilms of *Bdellovibrio - M. luteus* mixed cultures (right),

*Bdellovibrio - E. coli* and *M. luteus* mixed cultures (middle), and *Bdellovibrio*

*- E. coli* mixed cultures (left) were removed from growth medium and stained

with 0.1% crystal violet stain daily between day 1 to day 5.  The more dense the biofilm, the more blue is the coverslip. The biofilm density increases from day 1 to day 3, then decreases from day 4 to day 5. Compared to the biofilms without *Bdellovibrio*, there is a decrease in the staining color, which implies the predation of *Bdellovibrio* in the cultures.

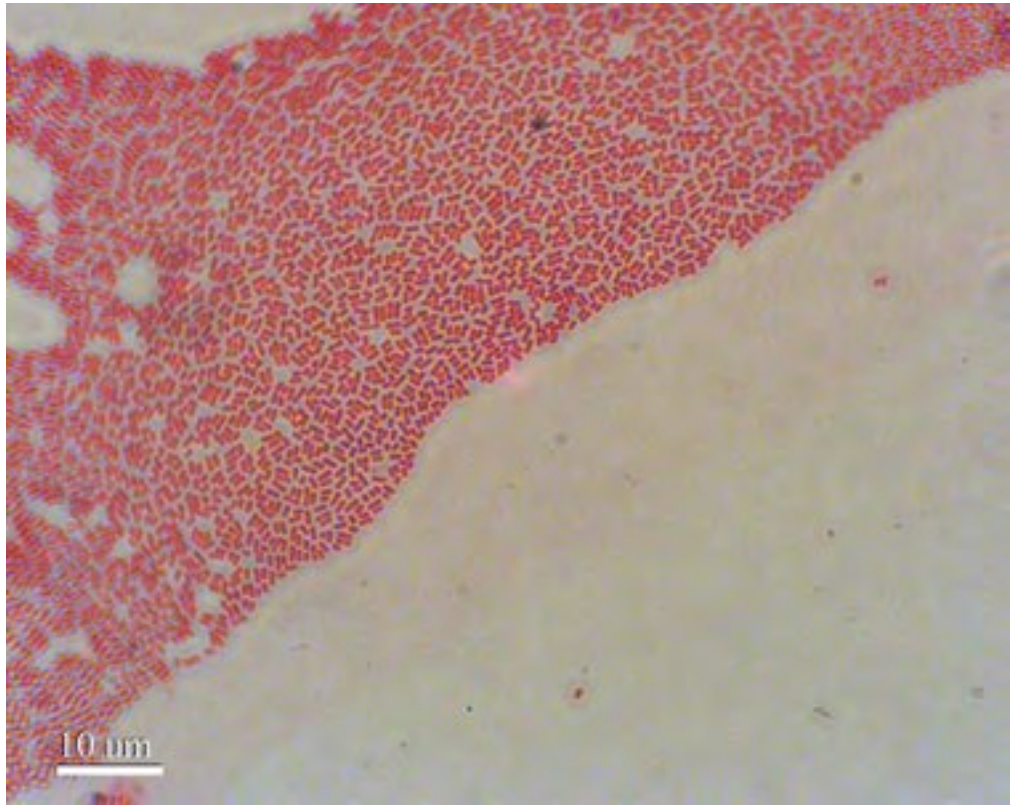

Figure 9: *E. coli* biofilm grows at the liquid-air interface, visualized using gram staining on day 1. The dense colony appeared at the interface between liquid growth medium and air, while almost no colonies appeared on the glass in air above the medium.

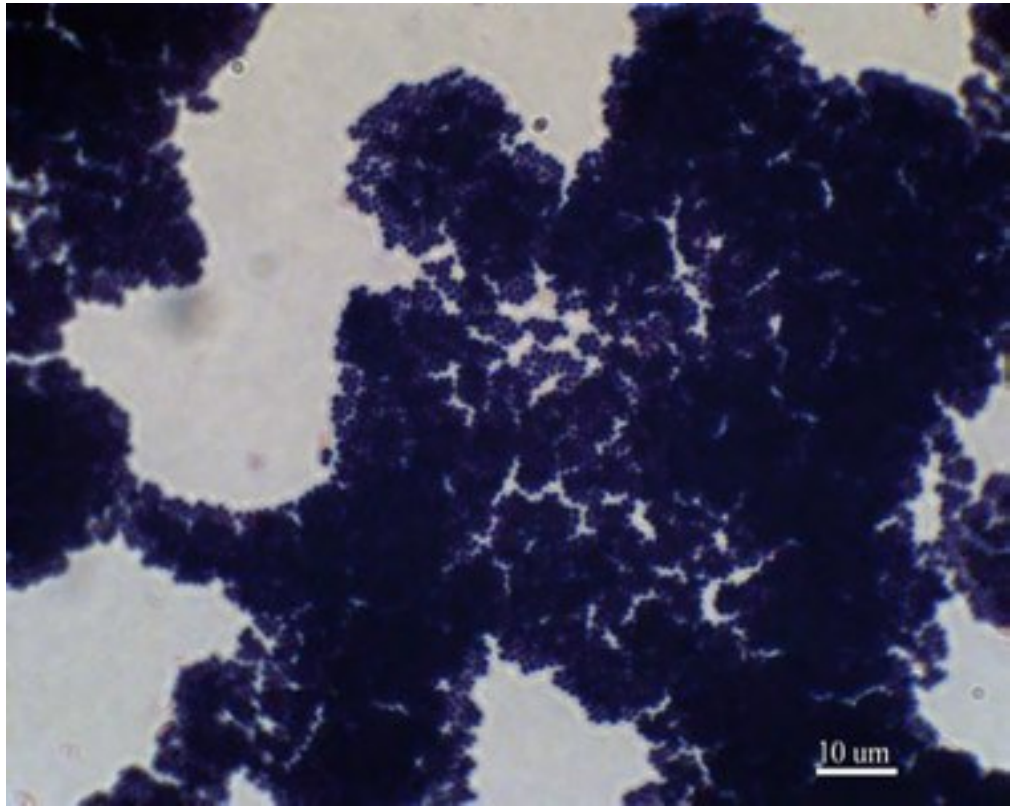

Figure 10: Light microscope image of a gram-stained *M. luteus* biofilm. *M. luteus* cells are round and they tend to form aggregates in a biofilm. The darker the stain color, the denser the population in the cluster. (Image taken by He Xu '12.)

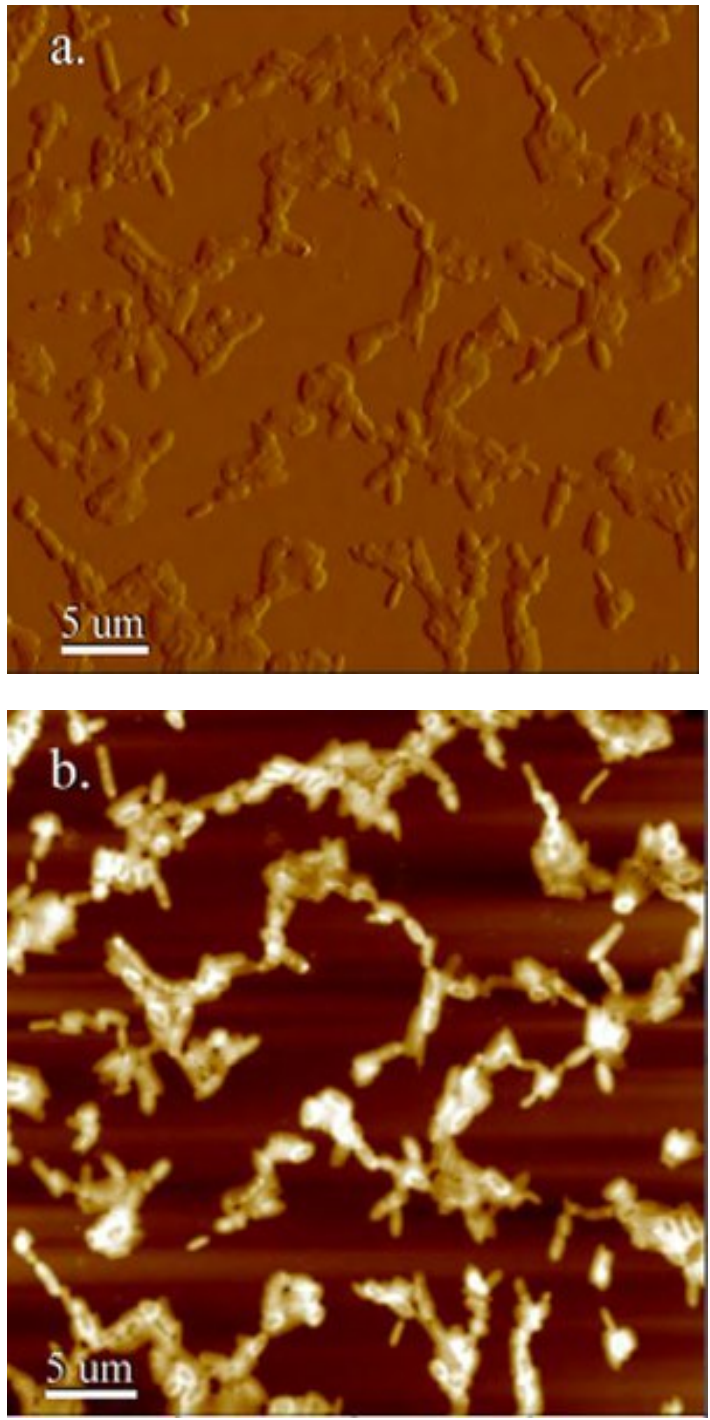

Figure 11: Atomic Force Microscope (AFM) images of *E. coli* biofilms. a)

Deflection image of *E. coli* spreading out and laying flat on the glass surface.

b) Corresponding height image of *E. coli*. The lighter the color, the higher the cells from the surface. *E. coli* cells mostly have the same height, vary between 200 nm to 400 nm (Images taken by He Xu '12).

Figure 12: Scanning Electron Microscopy (SEM) image of *E. coli* biofilms. Generally *E. coli* prefer to grow horizontally across the surface in a single layer as shown here. Only rarely do they grow on top of each other (data not shown).

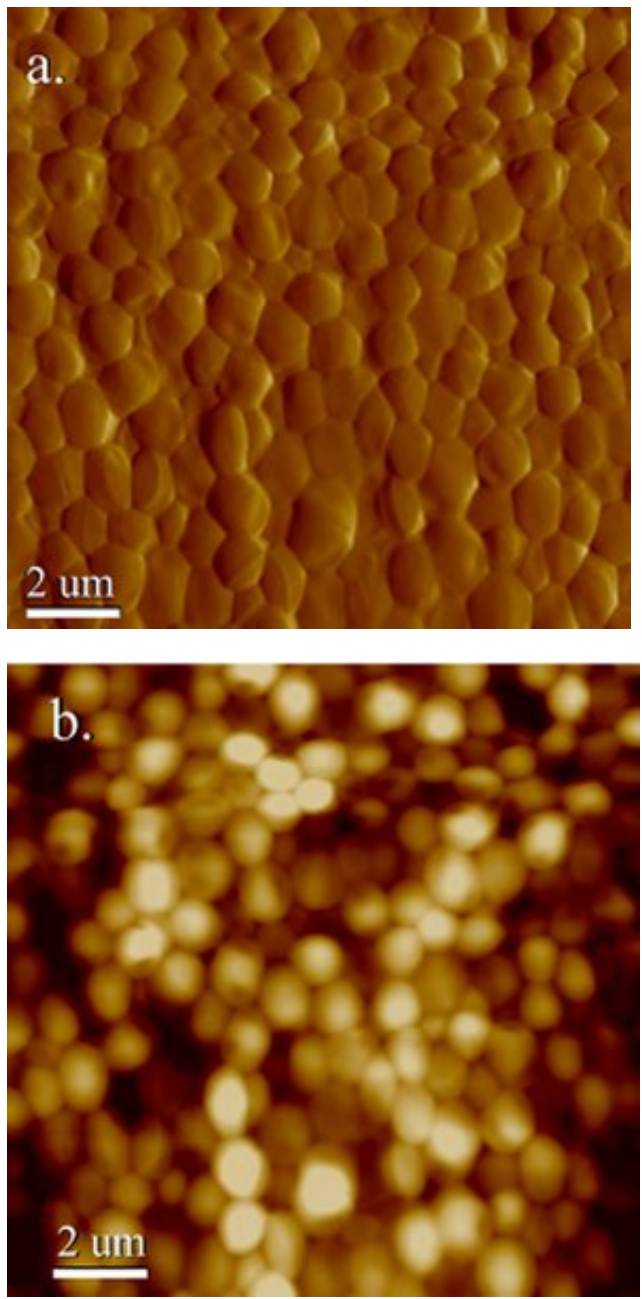

Figure 13: AFM images of *M. luteus* biofilms. The image is captured at 15 x

15μm scan size. (a) Deflection Image of *M. luteus* cells with a round, grape-like shape and a smooth texture. (b) Corresponding height image of *M. luteus*.

The lighter the color, the higher the cells from the surface. In this particular image, the difference in heights is ~ 800nm.

Figure 14: SEM image of a *M. luteus* cell cluster in a biofilm. *M. luteus* cells have a round, grape-like shape and a smooth texture. These *M. luteus* clusters form in a columnar fashion.

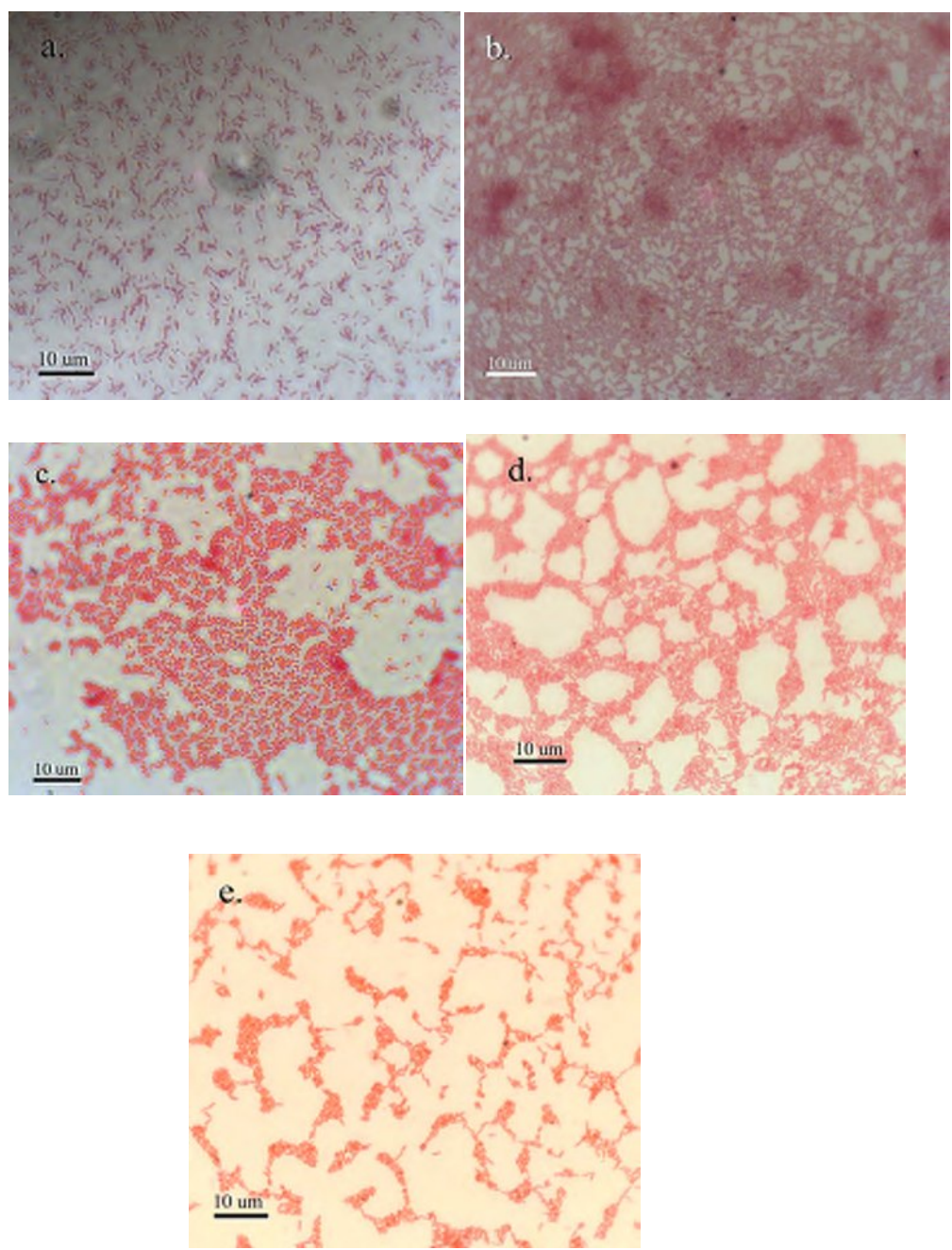

Figure 15: Light microscope images of gram-stained *E. coli* biofilm cells from

day 1 to day 5. (a), (b), (c), (d) and (e) are images of *E. coli* on day 1,2,3,4 and

5 respectively. The *E. coli* are more crowded on the first 3 days, and then start to decrease on day 4 and 5.

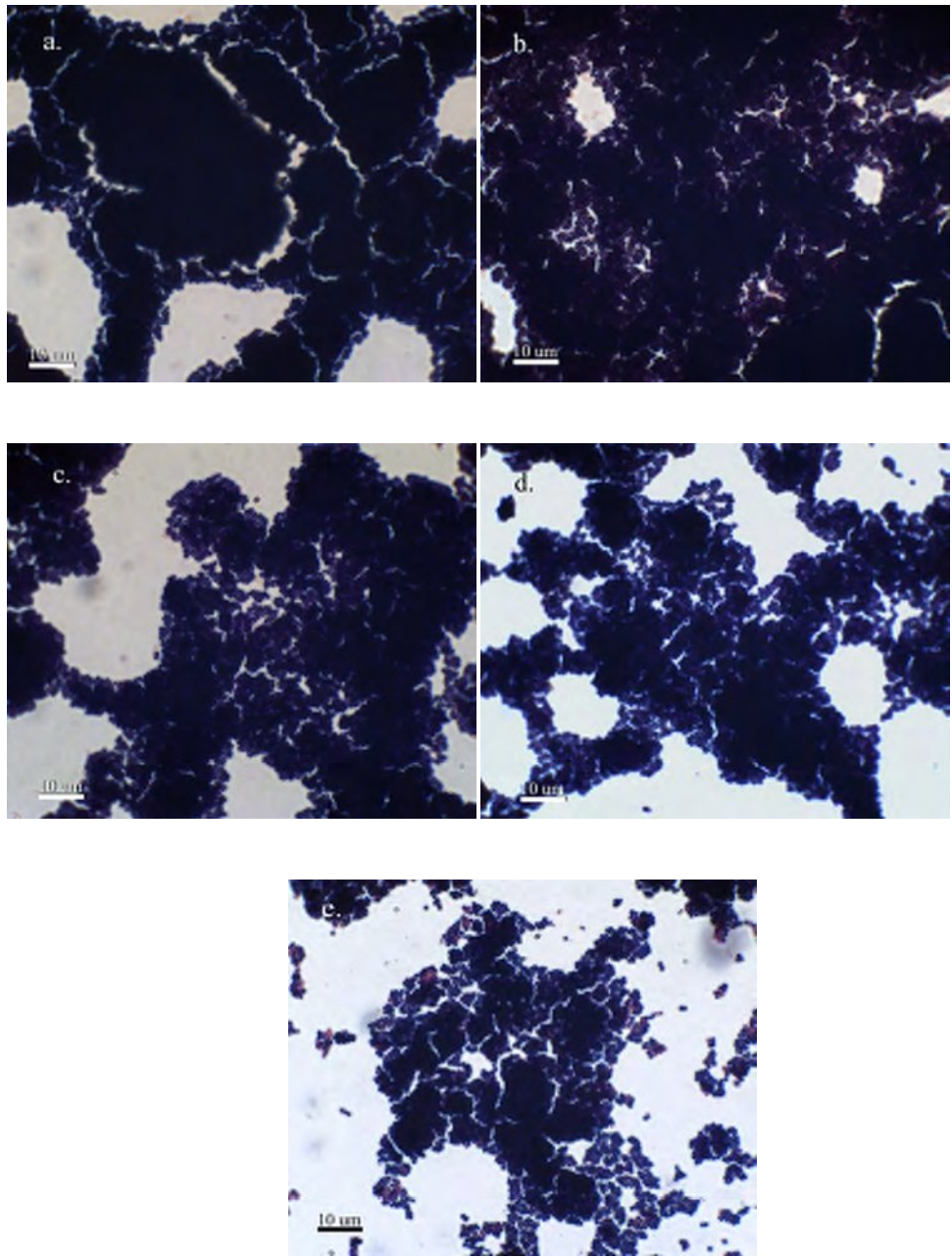

Figure 16: Light microscopic images of gram-stained *M. luteus* biofilm cells from day 1 to day 5. (a), (b), (c), (d) and (e) are images of *M. luteus* on day 1,2,3,4 and 5 respectively. The *M. luteus* are more crowded on the first 3 days, and then start to decrease on day 4 and 5. (Images taken by He Xu '12).

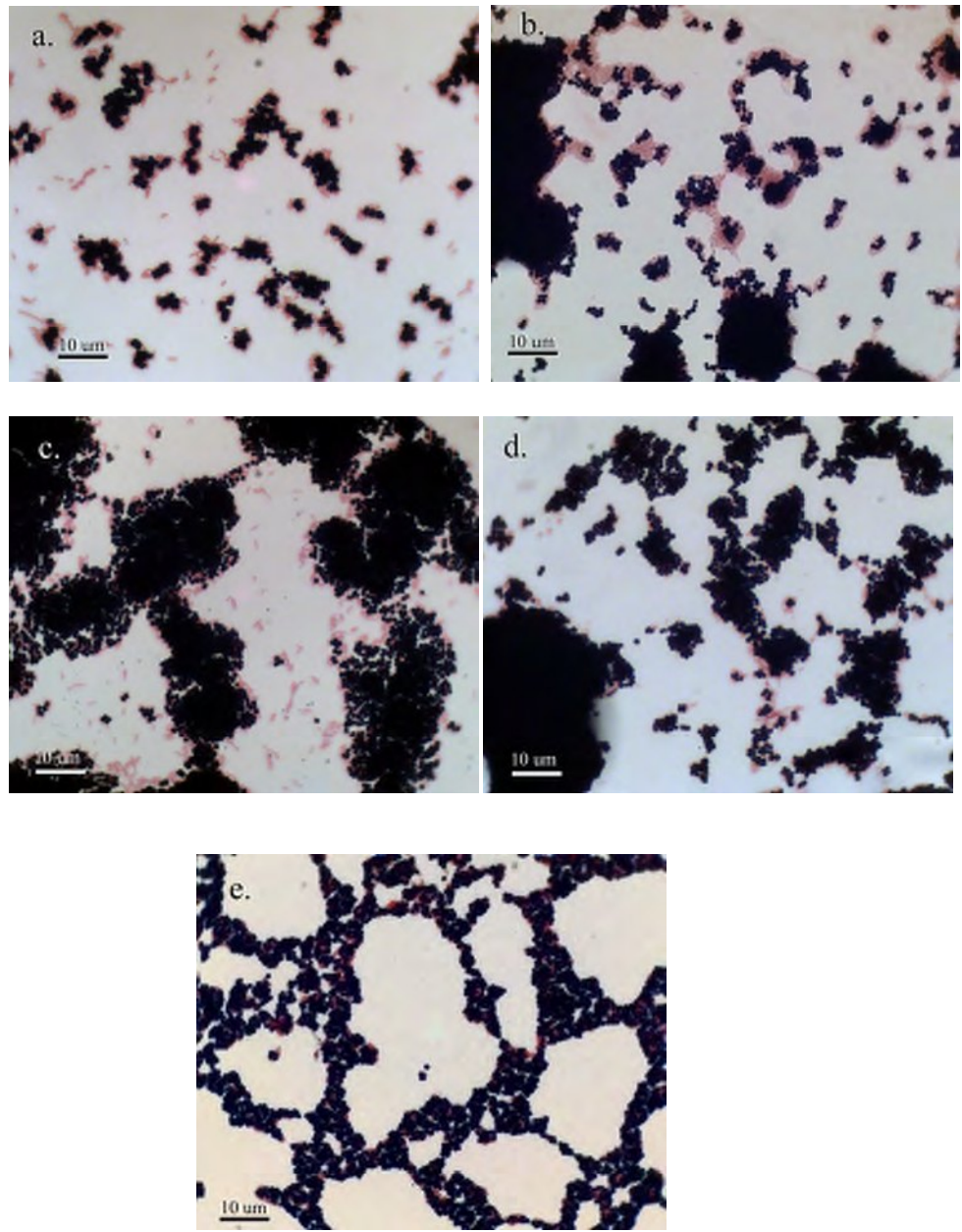

Figure 17: Light microscope images of gram-stained *E. coli* (pink) and *M. luteus* (purple) cells in mixed *E. coli* - *M. luteus* biofilms from day 1 to day 5. (a), (b), (c), (d) and (e) are images of the mixed biofilms on day 1,2,3,4 and 5 respectively. *M. luteus* grows in aggregates, often on top of the *E. coli*

monolayers.  *E. coli* (pink stained cells) can be seen on the first 3 days, but become more difficult to find on day 4 and 5, possibly because the *M. luteus* have grown to completely cover them.   There was no sign of the change in the number of *M. luteus*.

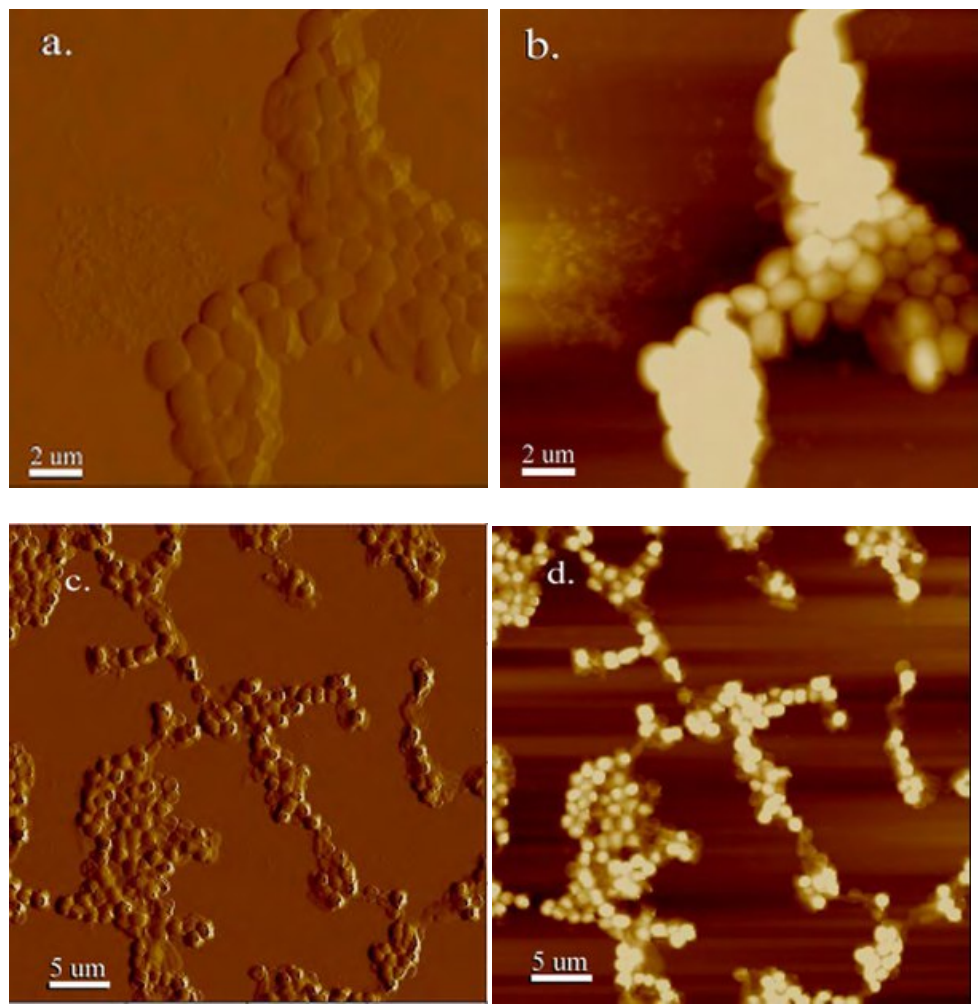

Figure 18: AFM images of mixed *E. coli* - *M. luteus* biofilms. (a) and (c) are deflection images of *M. luteus* cells with a round, grape-like shape laying on top of the single sheet oblong *E. coli*. EPS is secreted to allow bacteria to form a robust biofilm. (b) and (d) are corresponding height images of *E. coli* - *M. luteus* biofilms. There is a height difference between *M. luteus* cells and *E. coli* cells. Bright-lit *M. luteus* cells are 800nm – 1.5µm in height while dimmer *E. coli* cells are 200nm – 400nm tall above the surface. (Image 18 c and d taken by He Xu '12).

Figure 19: SEM images of mixed *E.coli* and *M.luteus* biofilms at 10 µm and

2.0 µm respectively. *M. luteus* cells have a round, grape-like shape and a smooth texture. These *M. luteus* cells form in a columnar fashion and appear to grow on top of rod-shaped *E. coli*.  *E.coli* are mostly at the bottom of *M.luteus* (top picture). Occasionally, there is a well-mixed community of *E.coli* and *M. luteus* where *E.coli* mingles with the *M.luteus* and sometimes stay on top (bottom picture). EPS secretion which appears in white strings allow robust biofilm formation.

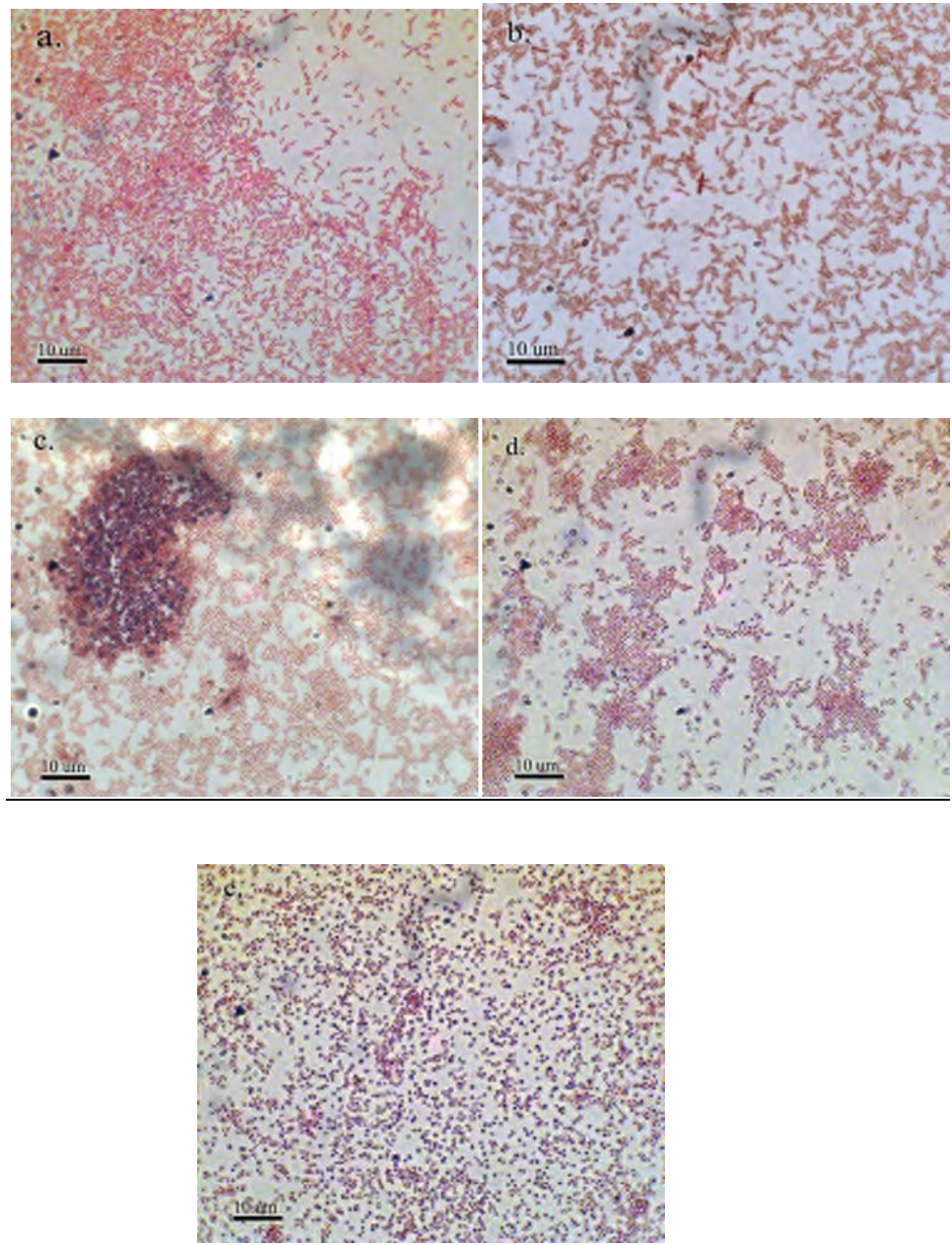

Figure 20: Light microscope images of gram-stained *E. coli* cells in mixed

*Bdellovibrio - E. coli* biofilms from day 1 to day 5. (a), (b), (c), (d) and (e) are images of the mixed biofilms on day 1,2,3,4 and 5 respectively. The pink and oblong *E. coli* grows in a monolayer. There is a slight increase in *E. coli* cell density on the first 2 days (a and b), and a dramatic decrease during later days (c, d and e).

Figure 21: SEM images of *Bdellovibrio* and *E. coli* in mixed *Bdellovibrio – E. coli* biofilms. The long, white string is the EPS secretion of bacteria to form a biofilm. *Bdellovibrio* and *E. coli* tend to lay flat on the surface. *Bdellovibrio* is approximately ~ 0.5µm longwhile *E. coli* is two times bigger. The arrow indicates a bdellovibrio attaching to an *E. coli* cell surface.

Figure 22: SEM image of *Bdellovibrio, E. coli,* and bdelloplasts in *Bdellovibrio – E. coli* biofilms. Bdelloplasts (arrows) are round and wrinkled in shape with a bdellovibrio growing inside. Due to the alcohol treatment before imaging, these two bdelloplasts do not appear as round as they are supposed to be. *Bdellovibrio* are small around $0.5\mu m$, and *E.coli* are bigger (around $\sim 1\mu m$). EPS formation is observed in long white strings allowing the formation of a biofilm.

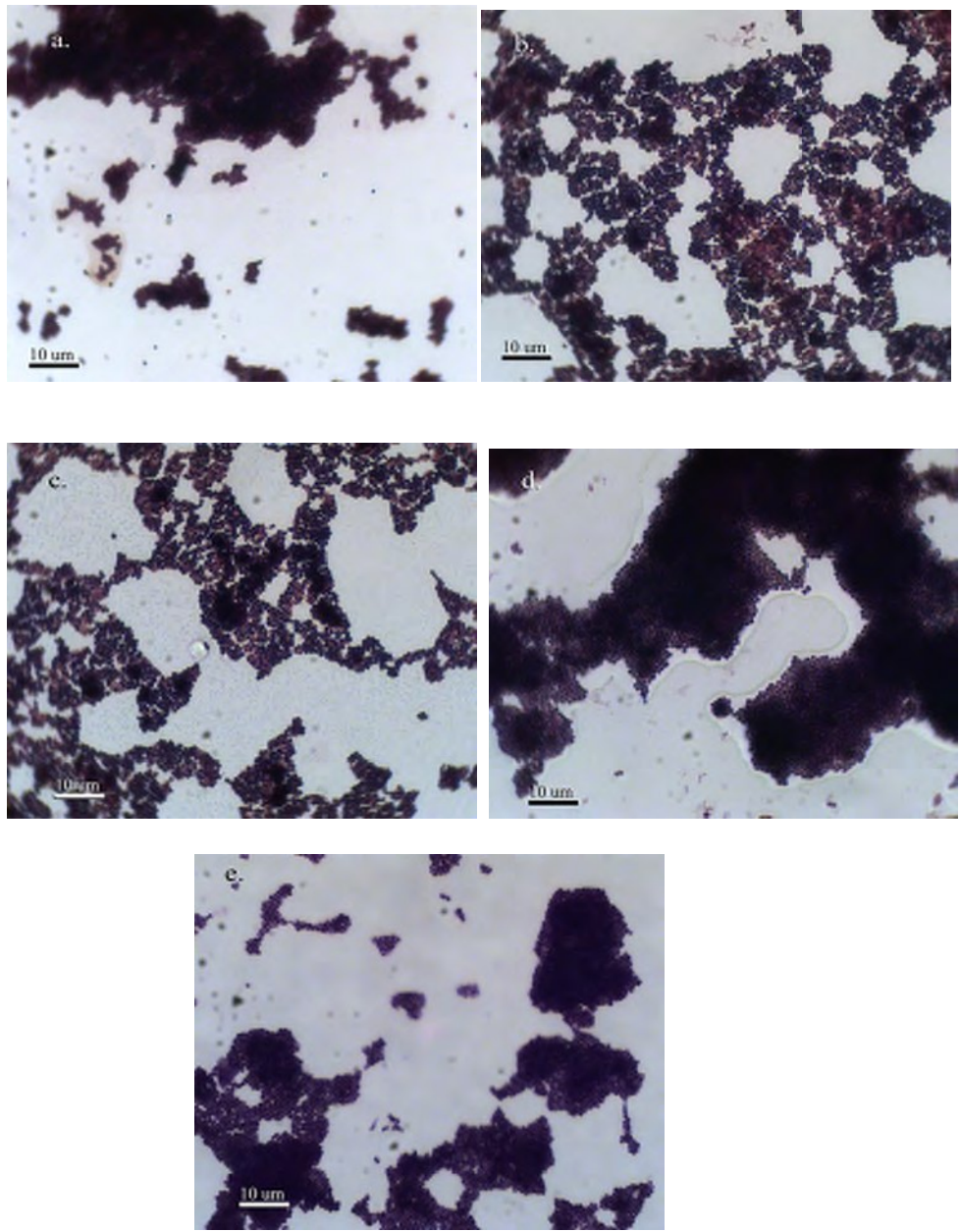

Figure 23: Light microscope images of gram-stained *M. luteus* (purple) cells
in mixed *Bdellovibrio - M. luteus* biofilms from day 1 to day 5. (a), (b), (c),
(d) and (e) are images of the mixed biofilms on day 1,2,3,4 and 5 respectively.
*M. luteus* grows in aggregates. There is a subtle decrease in *M. luteus* cell
density, however, the change is not clear.

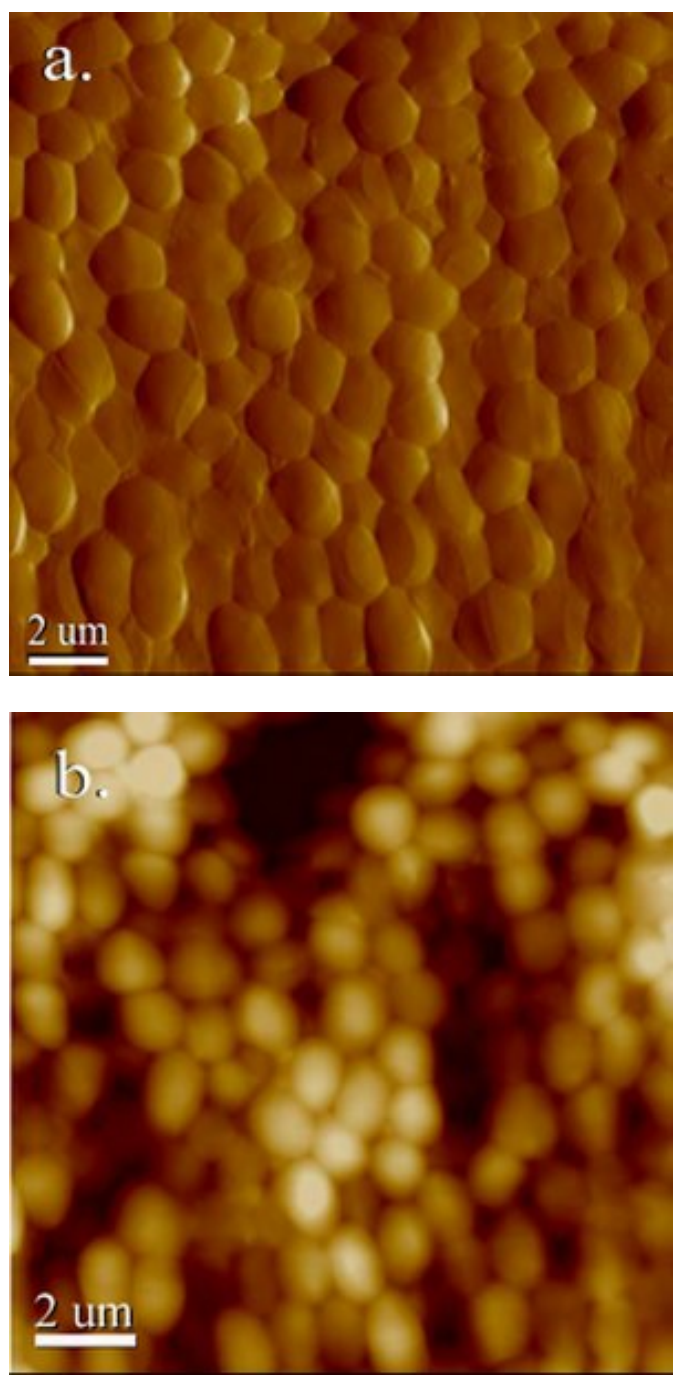

Figure 24: AFM images of *M. luteus* in *Bdellovibrio - M. luteus* biofilms with the scan size of 15 μm. The deflection image (a) shows the *M. luteus* round shape and smooth surface, while the height image (b) show the height variability between the cells. *M. luteus* cells can adhere on top of each other up to 800 nm.

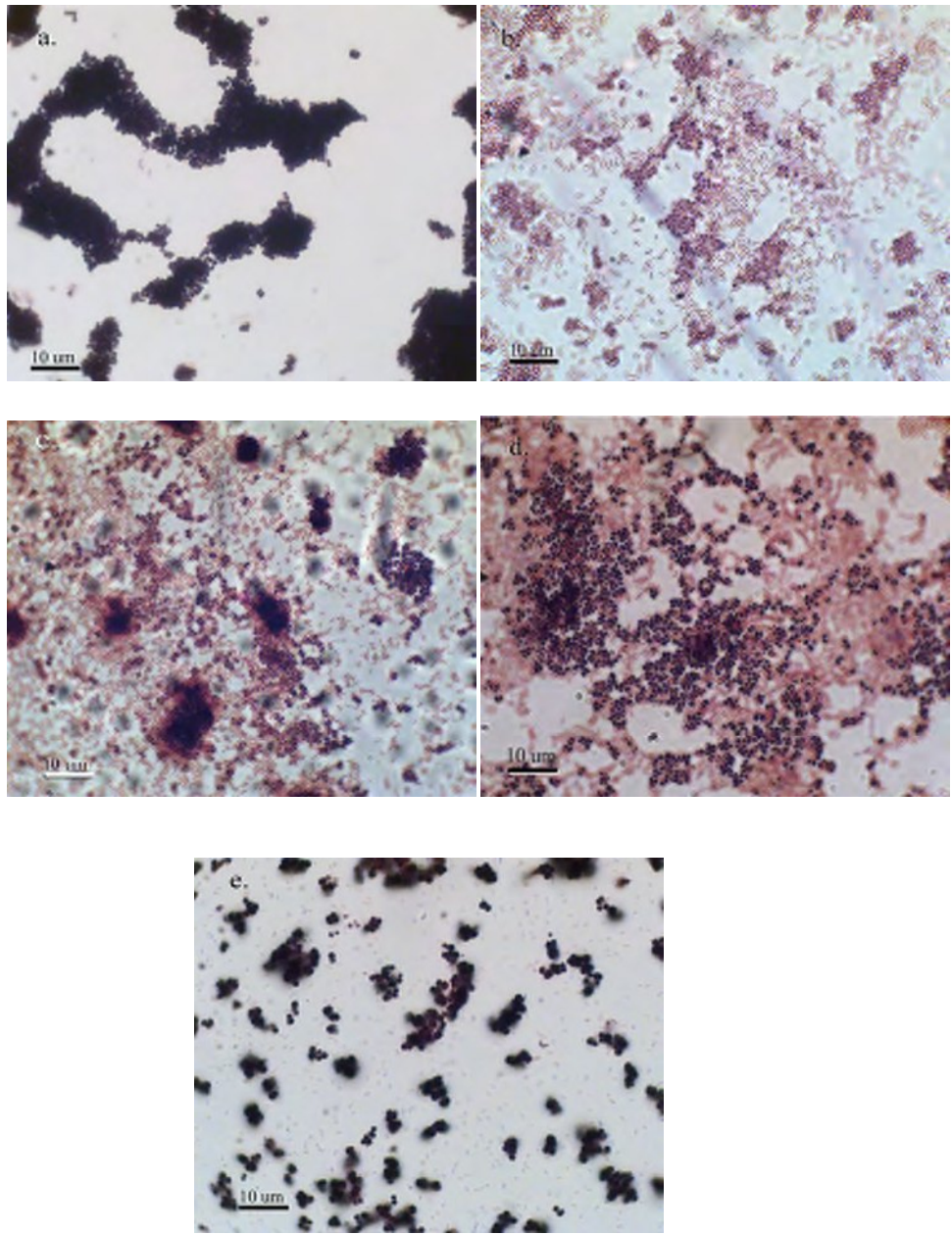

Figure 25: Light microscope images of *M. luteus* and *E. coli* cells in a mixed *Bdellovibrio – E. coli – M. luteus* biofilm from day 1 to day 5. The number of *E. coli* decreases over time. *M. luteus* cells are found to be less round, pack in units of two or four, and get darker over time.

Figure 26: SEM image of *E. coli and M. luteus* cells in mixed *Bdellovibrio –
E. coli – M. luteus* biofilms. A big portion of the biofilm has been eliminated.
Some *M. luteus* cells are found to be smaller and less round than others
(arrow). They also appear brighter in SEM image, indicating the possible
phenotype change due to the crowded environment.

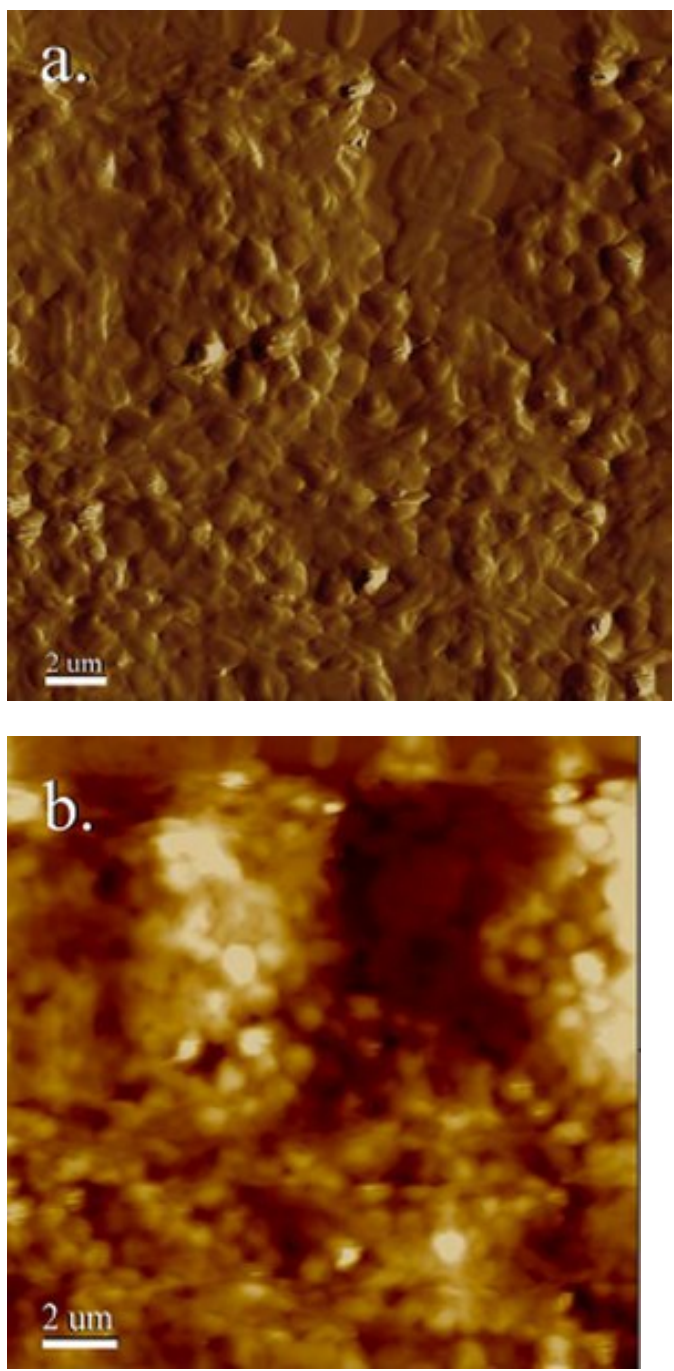

Figure 27: AFM images of mixed *M. luteus* and *E. coli* in *Bdellovibrio – E.*

*coli – M. luteus* biofilm at 20 μm scan size. The *E. coli* cells can be seen at the edges of *M. luteus*. *M. luteus* appears in round shape but its surface is not as smooth as usual due to the crowd and division process. (b) The height image shows the dark brown *E. coli* cells are 200 nm – 400 nm high, and some *M. luteus* appear in bright yellow and white color can be as high as 1.5 μm.

(Image taken by He Xu '12)

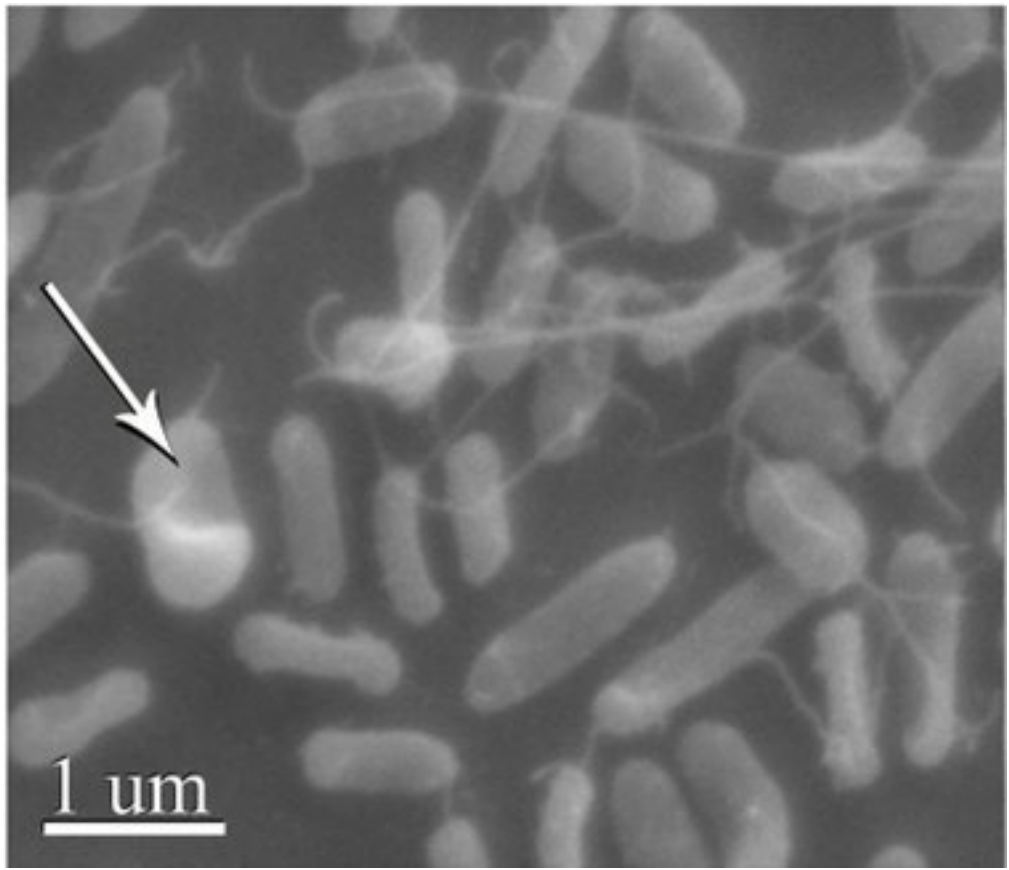

Figure 28: SEM image of a bdelloplast among *E. coli* cells in a mixed

*Bdellovibrio – E. coli – M. luteus* biofilm. The characteristic bdelloplast is less than 1μm in diameter.

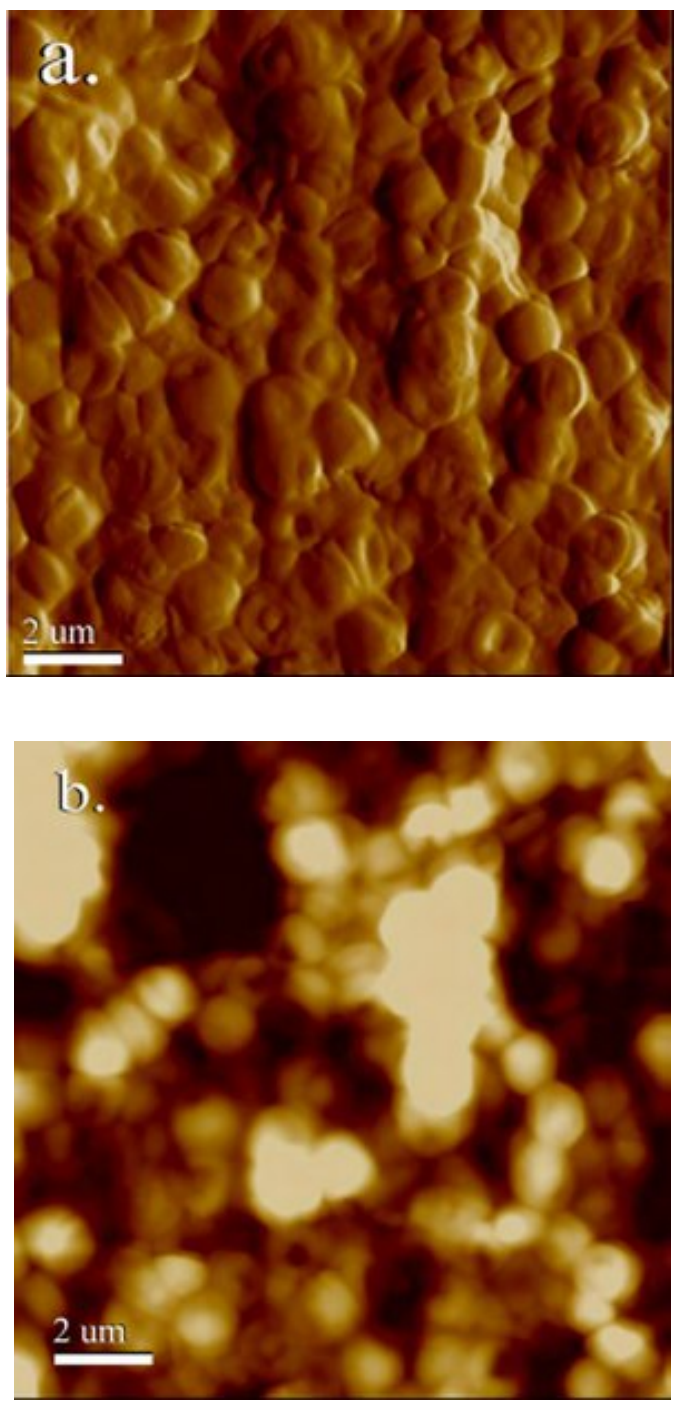

Figure 29: AFM images of *M. luteus* cells and bdelloplasts in mixed *Bdellovibrio – E. coli – M. luteus* biofilms. The differences between *M. luteus*

and bdelloplasts are not clear in this image due to the similarity in shape. (a) The deflection image shows a smooth shape *M. luteus* and a wrinkle shape of *Bdellovibrio.* (b) The height image shows that the cells can be adhere on top of each other up to 2.0μm away from the surface.

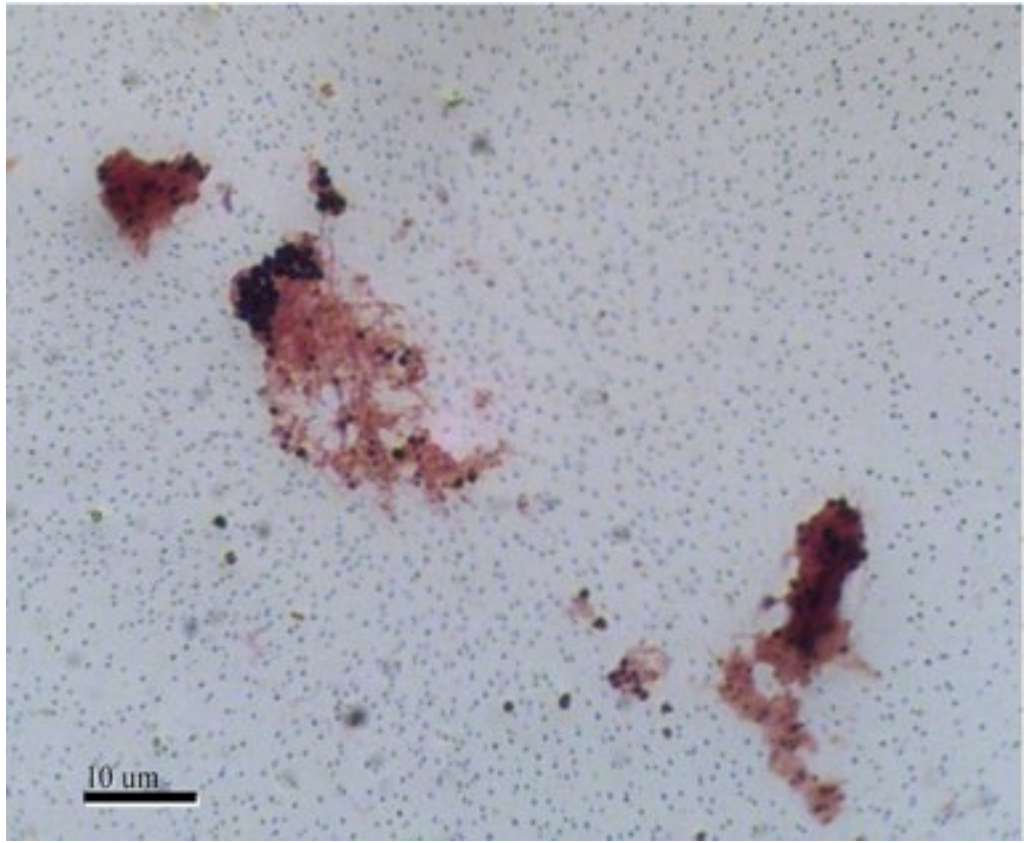

Figure 30: Light microscope image of gram-stained *M. luteus* and *E. coli* cells in a mixed *Bdellovibrio – E. coli – M. luteus* biofilm. Parts of *M. luteus* cell clusters fall off exposing *E. coli* foundation underneath. An empty space in the surrounding area indicate the elimination of *E. coli* cells in this biofilm.

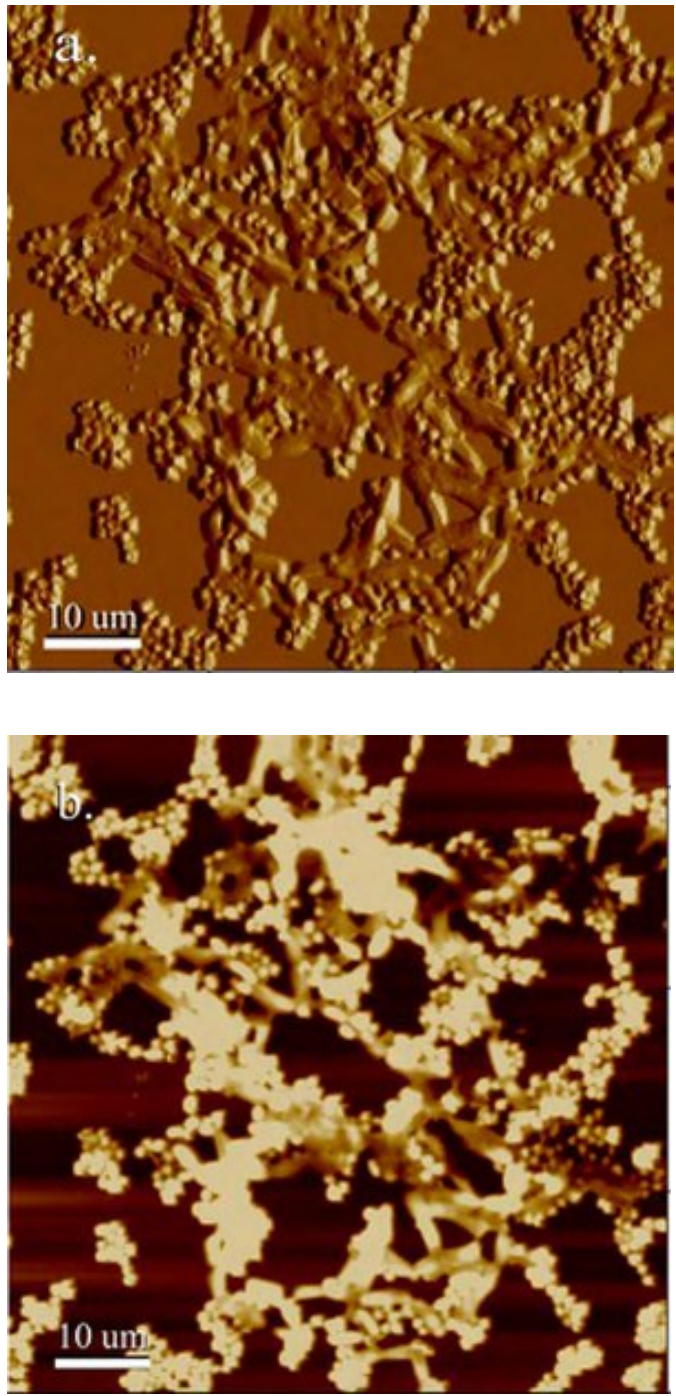

Figure 31: AFM images of *M. luteus* and *E. coli* cells in a mixed *Bdellovibrio –*

*E. coli – M. luteus* biofilm with the scan size of 60 x 60 μm. The deflection

image (a) shows *M. luteus* still clustered, but there are some parts that look like the cells are removed, thus revealing the *E. coli* foundation underneath. (b) The height image contains a bright lit color of *M. luteus* cells and dimmer color of *E. coli* cells. *M. luteus* cells in some area can be as high as 3.5 μm to 4.0 μm away from the surface.

Figure 32: SEM image of *E. coli, M. luteus* and *Bdellovibrio* in mixed *Bdellovibrio – E. coli – M. luteus* biofilms. In the *M. luteus* clusters, some portions fall off to expose *E. coli* cells underneath. By taking a closer look, we see that there are bdellovibrios (arrows) in the exposed, flattened *E.coli* portion.

# CHAPTER 4: DISCUSSION

## I. Population dynamics of *E. coli* and *M. luteus* with and without *Bdellovibrio* investigated with cell counting

The cell density of *E. coli* and *M. luteus* with and without *Bdellovibrio* was examined using cell counting technique.

### a. E. coli cell density

The availability of nutrients and oxygen is important for cell survival. Indeed, without competition for nutrition and oxygen, *E. coli* alone reaches the highest population. When *M. luteus* cells are added into *E. coli* biofilms, the *E. coli* population in the first two days is less than that without *M. luteus* (Figure 4). The crowded cells in the environment lead to the lack of nutrients and oxygen, thus reducing the cell density. During later days, the number of *E. coli* cells has a subtle decrease, which can be explained by the breaking down of *M. luteus* cell debris to provide extra nutrients for *E. coli*. This significant nutrition supply helps minimize the decrease in *E. coli* cell density after its preliminary growth, and supports a plateau in the growth of bacteria over the course of 5 days.

In addition to the lack of nutrition and oxygen, the presence of the predator *Bdellovibrio* further diminishes *E. coli* density. The predation by bdellovibrios speeds up the death of *E. coli* cells, and accounts for the sharper decline of *E. coli* cell density in mixed *Bdellovibrio – E. coli* cultures. *Bdellovibrio* predation might also explain the lower growth of *E. coli* cell counts on the initial days compared to the growth of *E. coli* biofilms without *Bdellovibrio*. In our study, the declining phase of *E. coli* cells with the presence of bdellovibrios started after 48-72 hours. This growth trend agrees with previous results conducted by this group (Nunez *et al.,* 2005). In Nunez *et al.* study, most *E. coli* cells were consumed starting after 48-72 hours of initial growth. These studies indicate that there is a lag phase in *Bdellovibrio* growth due to the slower life cycle compared to *E. coli* (Ruby *et al.,* 1991).

In mixed biofilms of *Bdellovibrio – E. coli – M. luteus,* our results provide evidence that *E. coli* growth is more restricted in a mixed biofilm of *Bdellovibrio – E. coli – M. luteus.* While in mixed *E. coli – Bdellovibrio* biofilms, the cell density of *E. coli* on day 5 ends roughly at the same level as the starting point on day 0, in the three-way mixed biofilm, the *E. coli* ending point is even lower than the starting level. Hobley *et al.,* (2006) also observed that the presence of gram-positive bacterium, *Bacillus subtilis* enhances the predation of *Bdellovibrio* on planktonic *E. coli* cells. We hypothesized that with the presence of *M. luteus* as a decoy in the biofilm, the biofilm becomes more crowded, thus increase the resource scarcity and *Bdellovibrio's* predation. *M. luteus* growing on top of *E. coli* in the cultures of *Bdellovibrio –
M. luteus – E. coli* does not help preventing the *E. coli* from *Bdellovibrio*
predation. *M. luteus* might secrete chemical signaling to attract *Bdellovibrio* to
come and attack the biofilms. This observation suggests that *Bdellovibrio* can
prey with high efficiency even when a gram-positive decoy is present.

### b. M. luteus cell density

Although the cell density of *M. luteus* is $10^2$ lower than *E. coli*
throughout, similar trends in population size are observed (Figure 5). Similar
to *E. coli* alone biofilms, biofilms of *M.luteus* alone reaches the highest
population density. It experiences a natural decline of population in the later
days of experiment period, which is caused by depletion of nutrition and
negative impact of crowding in the well.

When *E. coli* is added to the *M. luteus* biofilm, the population of *M.
luteus* decreases by 50%. Hobley and her colleagues suggested that when the
decoy and the prey interact in the same environment, they produce proteases
that can break down proteins (2006). These extra products generate abundant
nutrients for prey and decoy. *E. coli* can also do the same, and provide extra
nutrients for *M. luteus* cells, thus minimizing the drop of *M. luteus* in *E. coli –
M. luteus* biofilms. Therefore, *M. luteus* and *E. coli* can keep each other in
check to be well-maintained in a liquid culture. This might explain the minimal decrease in not only *E. coli* density but also *M. luteus* cell density in mixed *E. coli – M. luteus* cultures.

Interestingly, even though *M. luteus* is not consumed by *Bdellovibrio*, the presence of *Bdellovibrio* diminishes the *M. luteus* population. There is a small initial growth and the later significant decrease. *Bdellovibrio* is often described as an "obligate predator" (Stolp and Starr, 1963; Shilo, 1969; Rittenberg and Shilo, 1970; Thomashow and Rittenberg, 1978) and thus should not compete with *M. luteus* for nutrients. These data suggest that a life-style transition of *Bdellovibrio* from host - dependent (HD) to host - independent (HI) occurs while they are in the well liquid. Unlike HD *Bdellovibrio,* HI *Bdellovibrio* does not need a high concentration of prey cells. HI *Bdellovibrio* can be found in abundant nutrient environments, using nutrients in the environment to grow instead of prey cells. It can form biofilms on glass surface and grow well at various temperatures varying from 22˚C to 37˚C. At lower temperatures, the HI *Bdellovibrio* grows more slowly but has additional time to allow biofilm formation (Medina and Kadouri, 2009). In the absence of prey, our HD *Bdellovibrio* might have converted into HI *Bdellovibrio* during the course of 5 days and competed with *M. luteus* for nutrients, limiting *M. luteus*'s potential growth. This results in the relatively small initial expansion of *M. luteus* and its more prominent decline later.

When *E.coli* and *Bdellovibrio* are added to a *M. luteus* culture, competition from both *E. coli* and *M. luteus* for resources impedes the expansion of *M. luteus* even further. Our results here correlate with the studies of Hobley *et al.* (2006) In her study, the population of *Bacillus* decoys in a liquid culture of planktonic *Bdellovibrio* and *E. coli* shows a significant drop. *Bdellovibrio* is predator of gram-negative bacteria, thus it does not prey on *M. luteus* or *Bacillus* cells. The decline in *M. luteus* biofilm is not caused by *Bdellovibrio* predation directly. We hypothesized that there might be a switch between HD *Bdellovibrio* to an HI *Bdellovibrio*. The presence of HI *Bdellovibrio* would increase the scarcity in nutrient and oxygen supply and diminish *M. luteus* cells in a biofilm.

In this experiment, we do not directly measure bdellovibrio cells due to several constraints. Host-dependent (HD) *Bdellovibrio* is incapable of growing in a high nutrient medium that *E. coli* and *M. luteus* prefer. If HD *Bdellovibrio* can grow under such environments, it is most likely to convert to host-independent life style where nutrients are abundant. In this case, HI *Bdellovibrio* colonies appear to be yellow on an agar plate, and look like *M. luteus*. Due to these restrictions, we do not know the effect of *Bdellovibrio* on *M. luteus*. Since our experiments show that *Bdellovibrio* is able to hunt in the three-species biofilms of prey and decoy, we hypothesized that *M. luteus* does not discourage *Bdellovibrio* predation on *E. coli*.

## II. Biofilm formation is detected by crystal-violet staining

In this experiment, we studied *E. coli* and *M. luteus* cells in a simple biofilm model. This model represents well the biofilms in nature, as reflected by in the strong adhesion to the solid surface and the growth at an air-liquid interface. We monitored the biofilm density qualitatively by staining with crystal violet (Figure 6), which stains the biofilm on the coverslip deep purple. The portion of cells at the air-liquid interface shows the darkest stain, indicating a high concentration of cells in this area. This implies the abundance of organic molecules and oxygen in the air-liquid interface, which allows planktonic cells to form clusters and release EPS to form rigid biofilms. Williams *et al.,* (1995) suggest that the role of surfaces is important for attachment and growth of bdellovibrios and heterotrophic bacteria. Generally, we observed that there are more cells concentrated in the glass coverslips to form biofilms than the plastic surface (data not shown). This observation suggests the adhesion ability to the glass surface is higher than that of the plastic surface, making the glass coverslips an optimal environment for most bacterial biofilms in our study to attach and develop.

Biofilm density of *E. coli* and *M. luteus* with and without *Bdellovibrio* is consistent with our cell counting (Figure 7 and 8). Biofilms of *E. coli* alone and *M. luteus* alone exhibit the darkest color on the first two days, then the color decreases on the later days of the experiment. This observation fits the characteristics of a biofilm. The extent of biofilm accretion on surfaces is controlled by the amount of nutrient available for cell replication and EPS production. A rich nutrient environment is an optimal surface for bacteria to adhere, thus triggering biofilm formation through the secretion of EPS (Costerton *et al.,* 1995). However, bacteria do not form biofilms where the nutrients are lacking. They will leave the environment and convert back to the free-swimming life style (Williams *et al.,* 1995). With the same concepts, *E. coli* cells and *M. luteus* cells reach the highest population when the nutrients in the environment are optimal. During the latter days, the reduced nutrient availability, diminished oxygen concentrations due to the crowded environment, and possible release of potentially damaging metabolic by-products probably cause the drop in cell density.

In mixed *E. coli - M. luteus* biofilms, there is a reduction in the staining color of the biofilm compared to single species biofilms, indicating the decline of cell populations in a competitive environment. This result correlates with our cell counting results for the free-swimming cells of mixed *E. coli – M. luteus* culture.

All of the biofilms with *Bdellovibrio* involved show even more decrease in the staining color compared to biofilms without bdellovibrios. More cells were eliminated with the presence of bdellovibrios, implying the effect of bdellovibrio infection. The staining color of the *Bdellovibrio – M.*

*luteus – E. coli* biofilms is lighter than that of *Bdellovibrio – E. coli* and *Bdellovibrio – M. luteus* biofilms. The staining of the three-way mixed biofilms on day 5 is the clearest compared to the other days, indicating that with the presence of *M. luteus* decoy, *Bdellovibrio* predation is the most effective (Figure 8).

Biofilms offer good conditions for *Bdellovibrio*'s survival. It is suggested that in a biofilm, *Bdellovibrio* benefit from higher prey density improving the ease of searching and prey location (Williams *et al.,* 1995). Surfaces in aquatic environments provide the *Bdellovibrio* predator with essential nutrients for growth, which enhances the survival of *Bdellovibrio* under extreme environmental conditions. However, the fading in color indicating decrease in biofilm density is not as dramatic as we expected it to be. Perhaps, *Bdellovibrio* attack in a biofilm is not as vigorous as in the liquid culture. Kadouri and O' Toole, (2005) also suggest that bacteria in a biofilm have a greater survival when *Bdellovibrio* attack than bacteria in free – swimming phase even though *Bdellovibrio* is fully capable of hunting in a robust biofilm. Perhaps, bacteria in biofilms exhibit a phenotypic change that helps them be less susceptible to *Bdellovibrio* predation. The production of EPS prevents biofilms from negative effects from the environment such as antibiotics, bacteriophages, and chemicals; perhaps EPS also protects cells from *Bdellovibrio* predation (O' Toole *et al.,* 2000). It is also hypothesized that some population of bactera in the biofilm may be growing more slowly, be nutritionally deprived, or inducing a stress response, thus decreasing their susceptibility to *Bdellovibrio* attack (Mah and O' Toole, 2001). Nonetheless, no evidence that these factors slow *Bdellovibrio* down. *Bdellovibrio* eats UV-killed cells, and is not slowed down by capsules of polysaccharides (Varon and Shilo, 1968; Koval and Bayer, 1997). From a mathematical perspective, a 2D search is much easier than a 3D search. Nunez *et al.,* (2005) observed that *Bdellovibrio* hunts *E. coli* effectively at a surface. For these reasons, we believe that biofilms provide an excellent environment for *Bdellovibrio* to grow.

**III.  Bacterial interaction is revealed using light microscopy, scanning electron microscopy, and atomic force microscopy of biofilms**

In these experiments, light microscopy, SEM, and AFM were used to capture the interaction between bdellovibrios, *E. coli,* and *M. luteus.* Biofilms were imaged by light microscopy after gram-staining every 24 hours to see the overall change. AFM and SEM are performed to observe the biofilms on the nano-scale.

*a. Microscopic imaging of biofilms containing E. coli alone, M. luteus alone and E. coli – M. luteus Mixtures*

With gram-staining and light microscopy, *E. coli* cells appear pink and rod shaped, while *M. luteus* cells stain purple and have a round shape (Figures 9 and 10). In both cases, the darker the staining color, the denser the population. It is more common to find dark spots of staining color in *M. luteus* alone biofilms than that of *E. coli* alone biofilms, suggesting that *M. luteus* tend to form aggregates and adhere on top of each other, while *E. coli* are more spread out in a two-dimensional way on the glass surface. SEM and AFM images correlate with this observation. In the AFM height images, the lighter the color, the higher the cells from the surface. The darker the color, the flatter layer of the cells adhering to the surface. While *E. coli* biofilm is 200 – 400 nm in height, *M. luteus* alone biofilm can be as high as 700 – 850 nm (Figures 11 and 13). While *E. coli* cells in height AFM image have almost the same color, *M. luteus* cells show the variable in color, sometimes in bright lit and sometimes dimmer. This observation supports a monolayer structure of *E. coli* when they form biofilms, and *M. luteus* cells, on the other hand, cluster together in a columnar fashion.

When *E. coli* and *M. luteus* are mixed together in a biofilm, regardless of the presence of *Bdellovibrio*, the basic interaction between *E. coli* and *M. luteus* remains the same: *M. luteus* tend to form clusters and stay on top of the

*E. coli* monolayer. *E. coli* mostly adhere directly to the surface and are usually found in the edges of *M. luteus*. Therefore, when imaging the biofilm of *E. coli* and *M. luteus* by either AFM or SEM, it is hard to see *E. coli* cells in the biofilms. The biofilms seem to be overwhelmed by *M. luteus* covering the *E. coli* cells. The AFM height images of *E. coli* and *M. luteus* show a great difference in height between *E. coli* and *M. luteus* (Figure 18). While *E. coli* cells in a mixed *E. coli – M. luteus* biofilm do not change in height (200 – 400 nm) relative to *E. coli* alone, *M. luteus* cells are approximately 1.5 μm high above the surface. This difference is due to the *E. coli* laying underneath. In fact, the difference is equal to the *E. coli* height, demonstrating that *M. luteus* are actually adhere on top of *E. coli* cells, thus raising the height of a biofilm cluster.

Under SEM, *E. coli* are mostly found at the base of the biofilm, while *M. luteus* cells protrude from the surface in a columnar fashion. This finding suggests that *E. coli* has a better adhesion to the surface than *M. luteus* cells, but *M. luteus* has better adhesion to other cells. We propose that in a mixed biofilm of *E. coli – M. luteus, E. coli* acts as the biofilm colonizer and foundation. The adhesion ability of this strain of *E. coli* is probably better than that of *M. luteus*, since there are some *E. coli* sheets found alone, while *M. luteus* clumps are mostly found on top of the *E. coli*. This observation suggests the cooperation between two different types of bacteria in a multi-species biofilm. Williams *et al.,* (1995) showed that each type of bacteria has its unique adhesion mechanism so that when different kinds of bacteria cooperate to form a biofilm, positive characteristics would be selected. This response helps bacteria build functional biofilm consortia that protect and give advantages to both groups. In the case of mixed biofilms of *E. coli* and *M. luteus, E. coli* might conquer the surface first to give a better surface adhesion for *M. luteus* to grow. *M. luteus* with a better adhesion to the surface and *E. coli* with *M. luteus* growing on top will able to form a more robust biofilm and be protected from the negative impacts of the environment, such as toxic substances from antibacterial agents.

Most of our light microscopic images show an increase in cell density of *E. coli* and *M. luteus* on the first two days, followed by a decline in the population on the last days. This observation again correlates with our cell counting data and crystal violet results. The biofilm bacteria reached a stationary growth phase after 24 to 48 hours. The development of a biofilm is caused by the formation of EPS to facilitate the attachment of bacterial cells to the surface. In addition to the facilitation of initial attachment of bacteria to the surface and the formation of robust biofilm architecture, EPS protects biofilms from exogenous effect of the environment (O'Toole *et al*., 2000). Indeed, the secretion of EPS in *E. coli* and *M. luteus* biofilms is also captured by AFM and SEM in all biofilms. EPS became crowded on day 2 during biofilm maturation, indicating the rigidity of a biofilm.

The decline in cell density after 48 hours is due to the crowd of the bacteria in a mature biofilm, causing deficiencies in nutrients and oxygen. Bacteria need to escape from a biofilm when the nutrient supply is exhausted for a more favorable environment. At this time, polysaccharide lyase is secreted to facilitate the dispersal of cells (Alison *et al.,* 1998), and reduce the population of cells in a biofilm. Perhaps, *Bdellovibrio* enzymatic digestion and consumption of EPS mimic this lyase activity, signaling to cells that it is time to disperse during the last 4-5 days.

### b. Microscopic imaging of biofilms containing Bdellovibrio

Our microscopic images show a decrease in cell density in *Bdellovibrio* – containing biofilms compared to that of predator – free biofilms. With the addition of *Bdellovibrio* predation, the biofilm density reached the highest population after 48 – 72 hours and started to decrease from there. Besides the lack of nutrients and oxygen availability that causes dispersal phase, *Bdellovibrio* predation accounts for the drop in cell population. *Bdellovibrio* predator prefers a high prey density, thus exhibits its vigorous predation on when the biofilm size reaches the maximum growth. Due to the predation and nutrient deficiency, the biofilm size does not grow as much on the first 48 hours, and decreases dramatically on the latter days with the appearance of *Bdellovibrio*.

In *Bdellovibrio – E. coli* biofilms, *Bdellovibrio* and *E. coli* grow as a monolayer on top of the glass surface as an *E. coli* biofilm, but less densely. Due to the small size of bdellovibrios and bdelloplasts, and the low magnification of light microscope, bdellovibrios and bdelloplasts cannot be observed. Under SEM and AFM, round bdelloplasts with bdellovibrios growing inside are also found, indicating active predation is occuring to cause the death of *E. coli* cells. Our SEM images show a large portion of *E. coli* cells were eliminated. These images demonstrate the efficiency of *Bdellovibrio* infection on *E. coli* cells.

A mixed biofilm of *Bdellovibrio – M. luteus* continues to form clusters that grow in a columnar fashion. The gram-staining and light microscope pictures of *Bdellovibrio*-containing biofilms show a decrease in cell population compared to that of the predator-free biofilms. The decline in *M. luteus* biofilm is not caused by *Bdellovibrio* predation. *Bdellovibrio* preys only on gram-negative species, thus should not hunt on *M. luteus*. These results support a switch between HD *Bdellovibrio* and HI *Bdellovibrio*. The presence of HI *Bdellovibrio* would increase the scarcity in nutrient and oxygen supply and diminish *M. luteus* cells in a biofilm. The study conducted by Reiner and Shilo (1969) showed that HI *Bdellovibrio* was able to live in bacterial extracts of gram-positive *Bacillus subtilis, Bacillus cereus,* or *Micrococcus lysodeikticus,* indicating that abundant nuitrients encourage HI

*Bdellovibrio* to grow, even if those nutrients are not derived from gram-negative cells.

If the reduction in biofilm density is not caused by HI *Bdellovibrio*, we hypothesize that *Bdellovibrio* might secreting some biomolecules that prevent *M. luteus* from forming biofilms. These molecules can be toxic to *M. luteus* growth, or they might be able to prevent the EPS formation of *M. luteus* in a biofilm, so that the individual cells cannot form a permanent attachment to the submerged surface.

Although AFM has the magnification to capture *Bdellovibrio,* no *Bdellovibrio* were found but only *M. luteus.* Therefore, the deflection image of *M. luteus* and *Bdellovibrio* biofilm is similar to the *M. luteus* alone biofilm. However, the height AFM image shows increase in the height of *M. luteus* in *Bdellovibrio – M. luteus* biofilm compared to *M. luteus* growing alone. Notably, *M. luteus* height in *M. luteus – Bdellovibrio* biofilm is the same as *M. luteus – E. coli* biofilm ($1.5\mu m$). This observation indicates that *M. luteus* might grow on top of *Bdellovibrio*, preventing us from seeing the *Bdellovibrio* in a mixture.

Upon inspection of the three-way mixed *Bdellovibrio – E. coli – M. luteus* biofilms, although the light microscopic images do not provide quantitative results, we can clearly see qualitatively that *Bdellovibrio* helps eliminate the cell density of not only *E. coli* cells but also *M. luteus* cells.

Fewer *M. luteus* cells can be seen in this biofilm compared to the biofilms without *Bdellovibrio,* thus exposing more *E. coli* cells on the surface. Ocassionally, there are void spaces captured by light microscope, AFM and SEM in the three-way mixed biofilms. This effect is rarely seen in predator-free biofilms, indicating the drop in cell density in a both *E. coli* cells and *M. luteus* cells in a biofilm with the presence of *Bdellovibrio*. It is probably due to the lack of nutrients and oxygen, as well as the hunting of *Bdellovibrio*. *Bdellovibrio* might produce some organic molecules inhibiting the formation of *M. luteus* biofilms, probably by digesting the *M. luteus* EPS reducing adhesion to the surface.

Interestingly, in the three-way mixed of *Bdellovibrio – E. coli – M. luteus* biofilms*, M. luteus* cells are found to be less round, and pack in units of two or four. The staining color of *M. luteus* cells get darker over time. Interestingly, these tiny dot cells are not only found in light microscope but also  in SEM. The small, less round dark stain that comes in units of two or four can either be underdeveloped *M. luteus* due to the lack of nutrition or poorly stained bdelloplasts. If they are bdelloplasts, we wonder why these cells are bigger than the one we observed in *Bdellovibrio-E. coli* biofilm. But if they are underdeveloped *M. luteus*, we have not seen anything irregular under the AFM. These tiny cells can also be due to a phenotypic change of *M. luteus* during the predation of *Bdellovibrio* and *E. coli*. It is also possible that *Bdellovibrio* might have produced some small "controled" molecules that capture the growth of *M. luteus*. The substances produced by *Bdellovibrio* can also be toxic to *M. luteus* cells so that *M. luteus* might be more prone to senesence.

With the addition of *Bdellovibrio* in a *M. luteus* – *E. coli* biofilm, we observed that the biofilm becomes thicker. Occassionally, a biofilm can grow up to 2.0 µm to 4.0 µm in height (data not shown). However, *Bdellovibrio* is still able to succefully attack the thicker biofilm of *E. coli* and consume *E. coli* cells underneath thick biofilms of *M. luteus* cells. Kadouri *et al.,* (2005) suggest that *Bdellovibrio* is not restricted to surface of biofilm and *Bdellovibrio* can sucessfully attack a biofilm as thick as 3.0 µm. In fact, our AFM images show that *Bdellovibrio* is capable of attacking through a biofilm that has a height up to 3.5 µm - 4.0 µm, which is 10 times bigger than a bdellovibrio's size. We observed that the bonds between *E. coli* and *M. luteus* are weakened with the involvement of *Bdellovibrio*. *M. luteus* cell clusters on top of *E. coli* cells seem to fall off, exposing the *E. coli* foundation underneath to be susceptible for *Bdellovibrio* predation. *E. coli*'s texture is debilitated, losing a smooth surface. A small group of *M. luteus* spotted on top of the "impared" *E. coli* cells is found on both under light microscope, AFM and SEM. This observation suggests that the adhesion of *M. luteus* on *E. coli* cells wears off. With this mechanism, *Bdellovibrio* will be able to hunt on *E. coli* even when *E. coli* cells are hidden under high columnar clusters of *M. luteus* cells.

# CHAPTER 5: CONCLUSIONS

## I. Applications

Biofilms have negative effects on the environment, industry, and human and animal health. Biofilms can clog water pipes, compromise agricultural products and contaminate medical implants from contact lenses to artificial hearts. Unfortunately, bacteria in biofilms are resistant to chemicals, mechanical removal, bacteriophage, and antibiotic treatments. Stolp *et al.,* (1969) discovered that *Bdellovibrio bacteriovorus* consume gram-negative bacteria including *Salmonella, E. coli, Proteus, Pseudomonas, Serratia*. Nakamura *et al.,* (1970), observed that *Bdellovibrio* could reduce cell density of *Shigella flexneri* that caused conjunctivitis in rabbits, thus reduced the severity of the disease. *Bdellovibrio* was found to be unable to infect mammalian cells, making it a good candidate for a living antibiotic to treat infections (Sockett *et al.,* 2004*)*. This research demonstrates the capability of *Bdellovibrio* to prey upon and eliminate natural biofilms consisting of gram-negative and gram-positive bacteria, thus inspiring us to consider various ways that *Bdellovibrio* might be used productively for industrial, environmental and medical purposes.

## II. Conclusion

In this thesis, we examined the *Bdellovibrio* interaction in a dual-species biofilms of gram-negative *E. coli* prey and gram-positive *M. luteus* decoy. We were interested to see if *E. coli* cells become more, less or equally susceptible to *Bdellovibrio* attack with the addition of *M. luteus* in *Bdellovibrio – E. coli* biofilms, and whether *Bdellovibrio* predation might increase or decrease decoy biofilm formation.

Cell counting, gram staining, crystal violet staining, light microscope, SEM and AFM were powerful tools in investigating the cell density and the interaction between bacterial species in biofilms. We have successfully observed the cell population and the interaction biofilms containing *Bdellovibrio*, *M. luteus, E. coli* and various combinations thereof quantitatively and qualitatively from a macroscale to a microscale.

Our experiment has shown that *Bdellovibrio* consumes *E. coli* in a single-species biofilm and in a dual-species biofilms of *E. coli* prey and *M. luteus* decoy. The cell debris of *M. luteus* is broken down proteolytically to serve as a food supply to *E. coli* cells. The predation of *Bdellovibrio* on *E. coli* is not affected when *M. luteus* is present in the mixed biofilm. *E. coli* cells with the presence of both *Bdellovibrio* and *M. luteus* is depleted more that than without *M. luteus.* Thus, *M. luteus* serves as a crucial factor of balancing the environment in a mixed biofilm.

*Bdellovibrio* in a mixed biofilm of *Bdellovibrio – E. coli – M. luteus* also weakens the attachment of the decoy bacterium *M. luteus* to the solid surface, thus allowing *Bdellovibrio* to prey on the "edible" *E. coli* cells underneath. *Bdellovibrio* controls the biofilm-forming bacterial populations in the surrounding liquid environment. Our results hint the possible use of *Bdellovibrio* in eliminating harmful bacteria in biofilms in industry, environment, and medicine.

### III. Future work

For the future of our study, we would like to investigate the impact of nutrient availability using flow cell models, and look into the potential lifestyle alteration between host – dependent and host – independent *Bdellovibrio* in biofilms that involve decoys. We hope to gain a better understanding of *Bdellovibrio*'s nature for application purposes of eliminating biofilms in multi-species bacterial biofilms.